\documentclass[12pt,  draftclsnofoot,onecolumn,article,romanappendices]{IEEEtran}
\usepackage{setspace}

\usepackage{cite}
\ifCLASSINFOpdf
   \usepackage[pdftex]{graphicx}
\else
  \usepackage[dvips]{graphicx}
\fi
\usepackage[cmex10]{amsmath}
\usepackage{amssymb}
\usepackage{amsfonts}
\usepackage{amsthm}
\usepackage{array}
\usepackage{dsfont}
\usepackage{amsbsy}
\usepackage{epstopdf}
\usepackage{graphicx,color}
\usepackage{hyperref}

\newtheorem{proposition}{Proposition}
\newtheorem{remark}{Remark}

\long\def\symbolfootnote[#1]#2{\begingroup%
\def\thefootnote{\fnsymbol{footnote}}\footnote[#1]{#2}\endgroup}
\newtheorem{theorem}{Theorem}
\newtheorem{definition}{Definition}

\newcommand{\dv}{\mathbf} 
\newcommand{\mc}{\mathcal} 

\usepackage{algorithm} 
\usepackage{algpseudocode}

\algnewcommand{\Inputs}[1]{%
  \State \textbf{Inputs:}
  \Statex \hspace*{\algorithmicindent}\parbox[t]{.8\linewidth}{\raggedright #1}
}
\algnewcommand{\Initialize}[1]{%
  \State \textbf{initialization}
  \Statex \hspace*{\algorithmicindent}\parbox[t]{.95\linewidth}{\raggedright #1}
}

\begin{document}

\title{\mbox{Lossy Compression for Compute-and-Forward in} \mbox{Limited Backhaul Uplink Multicell Processing}}

\author{
  \IEEEauthorblockN{I\~naki Estella Aguerri \qquad \qquad Abdellatif Zaidi
  \\}

\thanks{ I\~naki Estella Aguerri and Abdellatif Zaidi are with Mathematical and Algorithmic Sciences Lab, 
France Research Center, Huawei Technologies Co. Ltd., 92100 Boulogne-Billancourt, France.
   This work was supported by Huawei Technologies Co. Ltd.     Email: \{\tt inaki.estella, abdellatif.zaidi\}{\tt @huawei.com.}
}
}

\maketitle
\begin{abstract} 
We study the transmission over a cloud radio access network in which multiple base stations (BS) are connected to a central processor (CP) via finite-capacity backhaul links. 
 We propose two lattice-based coding schemes. 
 In the first scheme,  the base stations decode linear combinations of the transmitted messages, in the spirit of compute-and-forward (CoF), but differs from it essentially in that the decoded equations are remapped to linear combinations of the channel input symbols, sent compressed in a lossy manner to the central processor, and are not required to be linearly independent. Also, by opposition to the standard CoF, an appropriate multi-user decoder is utilized to recover the sent messages. The second coding scheme generalizes the first one by also allowing, at each relay node, a joint compression of the decoded equation and the received signal. Both schemes apply in general, but are more suited for situations in which there are more users than base stations. We show that both schemes can outperform standard CoF and successive Wyner-Ziv schemes in certain regimes, and illustrate the gains through some numerical examples.
\end{abstract}


\IEEEpeerreviewmaketitle

\section{Introduction}


Together with fading, interference is one of the most limiting factors against high data rate communication in networks. The cloud radio access network (CRAN) architecture is a network topology in which base stations (BSs) are connected to a cloud-computing central processor (CP) via error-free finite capacity links. This architecture is generally seen as a possible means to alleviate the effect of interference, by enabling, at the central processor, some joint processing of the signals received by multiple base stations. Also, this network topology has some other appreciable features, such as low cost deployment of BSs and flexible network utilization.  

In a CRAN, each BS essentially acts as a relay node; and, so, it can implement classic relaying schemes such as amplify-and-forward, decode-and-forward, compress-and-forward \cite{Cover:1979} or more advanced forms such as compute-and-forward \cite{Nazer:IT:2011} and noisy network coding \cite{Lim:IT:2011:NoisyNetwork}. However, despite the ongoing effort, the optimal transmission strategy is still to be found.

Two particular schemes have attracted considerable attention for the uplink CRAN model, successive Wyner-Ziv \cite{Somekh:2007:IT,DelCoso:2009:TWir, Sanderovich:2009:IT, Hwan:2013:VT, ZhouYu:2013:JSAC,Park:2013:SPLett } and compute-and-forward \cite{Nazer:ISIT:2009,HongCaire:IT:2013}.  In the successive Wyner-Ziv scheme, every relay node forwards a compressed version of its received signal. The CP recovers the compressed output signals successively, and then utilizes them to decode the users' messages. In this scheme, the correlation between received signals at distinct BSs is exploited, via Wyner-Ziv source coding with side information at the decoder \cite{Wyner1978}, to reduce the backhaul requirements. Compute-and-forward coding, in its standard form, requires each relay to decode one or more equations (with integer-valued coefficients) that relate the users messages, and then send them (uncompressed) to the CP. The equations are required to be linearly independent so that, with enough of them, the CP can invert the linear system to recover the users' messages. The two approaches are combined appropriately in \cite{Estella:Allerton20015} to balance the amount of information decoded centrally and distributedly; and it is shown that the resulting scheme can outperform strictly the best of the aforementioned two schemes.

While compress-and-forward has been studied in several distinct settings \cite{Somekh:2007:IT,DelCoso:2009:TWir, Sanderovich:2009:IT, Hwan:2013:VT, ZhouYu:2013:JSAC,Park:2013:SPLett }, the performance of compute-and-forward for CRAN-type networks has been studied mostly in settings in which there are more relays than users. In such settings, it is enough that every relay node decodes one equation (assuming that the equations are linearly independent). In the more practical settings in which there are more users than relay nodes, straightforward variations of compute-and-forward can be made to apply, e.g., by requiring that at least some of the relay nodes compute and forward more than one equation each. However, this generally results in some performance degradation. For the  application and analysis of CoF for some related multiaccess models, the reader may refer, e.g., \cite{E-SZV13} and \cite{E-SZV14} and the references therein.

On the other hand, in CoF, the decoded equations at the BSs relate the users' messages, and are therefore correlated.  This redundancy at the users' message level has been exploited in \cite{Tan2016Assym}, and\cite{ExpCof:NazerCNC15}, to reduce the backhaul requirements by forwarding lossless compressed versions of the decoded equations to the CP.  In this work, we consider a different approach and propose two lattice-based coding schemes that utilize distributed \textit{lossy compression} to exploit the redundancy in the decoded equations at the users' input symbol level (as opposed to the users' messages in \cite{Tan2016Assym}). Both schemes apply in general CRAN type topologies, i.e., for an arbitrary number of users and relay nodes, but are better suited for the situations in which there are more users\mbox{ than relays.}

 In the first coding scheme,  to which we refer hereafter as ``quantized-compress-and-forward" (QCoF), each BS first decodes a single linear combination of the users' messages, in the spirit of CoF. However, instead of forwarding it as is, like in standard CoF and its variants, the equation is first remapped to one on the users' input symbols and then sent compressed to the central processor -- the compression is performed taking into account the side information that is available at the CP, i.e., through Wyner-Ziv compression. At the CP, the messages are decoded successively from the decompressed signals using a suitable multi-user decoder. The main advantages of QCoF over standard CoF and straightforward variants of it can be summarized as follows. First, as every relay node is required to compute only one equation irrespective to the number of users, the scheme is more suited comparatively to the typical situations in which there are more users than relays. Second, since the computed equations need not be linearly independent with each other, the relay nodes need not coordinate among them or through the CP, which makes it more suitable in practice. The main advantage of QCoF over the scheme successive Wyner-Ziv can, at a high level, be seen as that of denoising the relay's output before compressing it.
In the second coding scheme, to which we refer hereafter as ``jointly -quantized-compute-and-forward" (JQCoF), we generalize QCoF by allowing each relay node to compress not only the computed equation on the users' input signals but also its received signal, from which the equation has been computed. The compression is performed jointly, through multivariate Wyner-Ziv compression. This allows to trade-off appropriately the amount of pure information that is sent to the CP and the amount of output that is conveyed compressed to the CP.
 We show that the scheme QCoF outperforms the standard CoF and successive Wyner-Ziv in certain regimes, which we illustrate through some numerical examples; and the scheme JQCoF  strictly outperforms all other schemes in general. 


	

\begin{figure}
\centering
\includegraphics[width=0.495\textwidth]{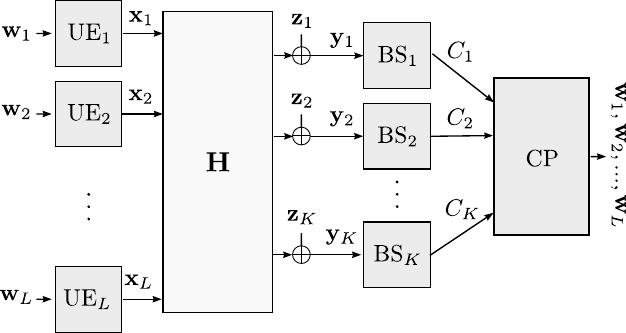}
\caption{The uplink of a Gaussian Cloud Radio Access Network (CRAN) with error-free finite-capacity backhaul links.} \label{fig:Schm}
\vspace{-5mm}
\end{figure}

\subsection{Outline and Notation}

The rest of the paper is organized as follows. In Section \ref{sec:System}, we describe the system model, and recall some basics on lattice coding.  In Section \ref{sec:QCoF}, we analyze the scheme QCoF; and in Section \ref{sec:JQCoF} we analyze the scheme JQCoF. Finally, Section \ref{sec:WynerModel} provides some numerical examples.

Throughout the paper we use the following notations. Lower case letters are used to denote scalars, e.g., $x$; upper case letters are used for random variables, e.g., $X$, boldface lower case letters  are used to denote vectors, e.g., $\dv x$; boldface upper case letters are used to denote matrices, e.g., $\dv X$; calligraphic letters are used to denote sets, e.g., $\mathcal{X}$.  The cardinality of a set $\mc X$ is denoted by $|\mc X|$. We use the notation $\mathbb{E}_{X}[\cdot]$ to denote the expectation of random variable over $X$, or $\mathds{E}[\cdot]$ if it is clear from the context. For integers $i \leq j$, we define $[i:j]:=\{i,i+1,\hdots,j\}$.  
We denote by $\mathds{R}_+$ the positive real numbers, and by $\mathds{F}_p$ the finite field of size $p$, where $p$ is always assumed to be prime.  We denote the transpose of a a vector $\mathbf{x}$ by $\mathbf{x}^T$ similarly for matrices, e.g., $\mathbf{X}^{T}$, and by $\text{diag}(\cdot)$ the operator that given a vector $\mathbf{x}$ generates a diagonal matrix with the elements of $\mathbf{x}$ in its diagonal, or a block diagonal matrix if applied on a set of square matrices, e.g., $\text{diag}(\mathbf{X}_1,\ldots,\mathbf{X}_K)$. We denote by $[X]_{j,k}$ the element in row $j$ and column $k$ of matrix $\mathbf{X}$, and  by $\mathbf{I}_{n}$ the identity matrix of size $n$ and by $\log^+(\cdot)=\max\{0,\log(\cdot)\}$. Finally, throughout the paper, logarithms are taken to base $2$.


\section{System Model}\label{sec:System}

We consider an uplink cloud radio access network (CRAN) model in which $L$ single-antenna user equipments (UEs) communicate with a central processor (CP) through $K$ single-antenna relay base stations (BSs). The model is shown in Figure~\ref{fig:Schm}. Each BS is connected to the CP via an error-free finite-capacity backhaul link of capacity $C_k$.
UE $l$, $l=1,\hdots,L$, wants to transmit a message $\mathbf{w}_l\in \mathds{F}_p^{k_l}$ to the central processor (CP). The message $\mathbf{w}_l$ is assumed to be uniformly distributed over the prime-size finite field  $\mathds{F}_p^{k_l}$. The rate of each message is given by $R_l=k_l/n\log p$. We assume that the messages are zero padded to have a common length $k\triangleq\max_lk_l$.  
In order to transmit its message, UE $l$ uses an encoder $f_l: \mathds{F}_p^{k}\rightarrow \mathds{R}^{1\times n}$ to map the message $\mathbf{w}_l$ into a length-$n$ channel input sequence $\mathbf{x}_l=f_l(\mathbf{w}_l)\in\mathds{R}^{1\times n}$. The encoding is subjected to the following average power constraint,
\vspace{-2mm}
\begin{equation}\label{eq:powConst}
\frac{1}{n}\sum_{t=1}^n \mathrm{E}[|\mathbf{x}_l(t)|^2]\leq \mathrm{SNR},
\quad l=1,...,L.
\end{equation} 
\vspace{-1mm}
The channel output at BS $k$, $k=1\hdots,K$, is given by
\begin{IEEEeqnarray}{rCl}\label{eq:ChanModSimple}
\mathbf{y}_k=\mathbf{h}_k\mathbf{X}+\mathbf{z}_k, 
\end{IEEEeqnarray}
where $\mathbf{h}_k=[h_{k,1},...,h_{k,L}]$ is the vector of channel coefficients from the users to BS $k$, $h_{k,l}$ denotes the channel coefficient from user $l$ to between BS $k$; $\mathbf{X}=[\mathbf{x}_1^T,...,\mathbf{x}_L^T]^T\in\mathds{R}^{L \times n}$ and $\mathbf{z}_k$ is the length-$n$ additive ambient noise sequence at BS $k$, whose elements are assumed to be independent and identically distributed (i.i.d.) Gaussian random variables with zero mean and unit variance , i.e., $\mathbf{z}_{k}(t)\sim \mathcal{N}(0,1)$. In order to relay its information, BS $k$, $k=1,\hdots,K$, maps the channel output $\mathbf{y}_k$ into an index $J_k=f_k^r(\mathbf{y}_k)\in[1:2^{nC_k}]$ which is then transmitted on the pipe to the CP. The CP collects all the indices $\{J_1,\hdots,J_K\}$ and estimates the users' messages $\{\mathbf{w}_1,\hdots,\mathbf{w}_L\}$. Let $g: [1:2^{n(C_1+\cdots+C_K)}]\rightarrow\mathds{F}_p^{k_1+\cdots+k_L}$ denote the decoding function at the CP.


\begin{definition}
For given channels $\mathbf{h}_1,\ldots,\mathbf{h}_K$, signal-to-noise ratio $\mathrm{SNR}$, and finite-capacity links of rates $\{C_1,\hdots,C_K\}$, we say that a rate tuple $R_1,\ldots,R_L$ is achievable if, for any $\epsilon>0$, there exist a sequence of encoding functions $f_l$, $l=1,\ldots,L$, relaying functions $f_k^r$, $k=1,\ldots,K$ and a decoding function $g$, such that, for sufficiently long blocklength $n$, each user's message can be decoded by the CP at rate at least $R_k$ with vanishing probability of error, i.e.,
\begin{IEEEeqnarray}{rCl}
 \mathrm{Pr}\{(\mathbf{w}_1,\hdots, \mathbf{w}_L)\neq (\hat{\mathbf{w}}_1,\hdots, \hat{\mathbf{w}}_L)\}\leq \epsilon. 
 \end{IEEEeqnarray}
\end{definition}

In the rest of this paper, we develop two lattice-based coding schemes and analyze the rate tuples that they achieve. For the sake of simplicity, we  restrict to real-valued channels; the analysis extends straightforwardly to complex-valued channels.

\subsection{Basics on Lattice coding} 

The proposed schemes are based on nested lattice codes. To simplify the exposition later, we provide the following standard  definitions which can be found, e.g., in~\cite{CS88}. A lattice $\Lambda$ is a discrete subgroup of $\mathds{R}^n$  which is closed under reflection and addition. It is characterized by $\Lambda=\{\boldsymbol\lambda = \mathbf{G}\mathbf{c}: \mathbf{c}\in \mathds{Z}^n\}$, where $\mathbf{G}\in\mathds{R}^{n\times n}$ is the lattice generator matrix. We denote by $\mathcal{V}$ the fundamental Voronoi region of lattice $\Lambda$. Also, let, for $\dv x \in \mathbb{R}^n$,  $Q_{\Lambda}(\mathbf{x})$ be the nearest neighbor lattice point to $\dv x$, i.e.,  $Q_{\Lambda}(\mathbf{x}) = \arg\min_{\mathbf{t}\in\Lambda}\|\mathbf{x}-\mathbf{t}\|$. The modulo operation with respect to (w.r.t.) lattice $\Lambda$ is defined as $[\mathbf{x}]\mod \Lambda=\mathbf{x}-Q_{\Lambda}(\mathbf{x})$. The second moment of $\Lambda$ is given by
\begin{IEEEeqnarray}{rCl}
\sigma^{2}(\Lambda)\triangleq\frac{1}{n}\frac{1}{\mathrm{Vol}(\mathcal{V})} \int_{\mathbf{u}\in\mathcal{V}}\|\mathbf{u}\|^2d\mathbf{u},
\end{IEEEeqnarray}
with $\mathrm{Vol}(\mathcal{V})$ denoting the volume of $\mathcal{V}$. The normalized second moment of $\Lambda$ is defined as $G(\Lambda)\triangleq\sigma^2(\Lambda)/(\mathrm{Vol}(\mathcal{V}))^{2/n}$.

\noindent A lattice $\Lambda_c$ is said to be nested in a lattice $\Lambda_f$ if $\Lambda_c\subseteq\Lambda_f$. Lattice $\Lambda_c$ is referred to as the coarse lattice and $\Lambda_f$ is referred to as fine lattice. More generally a sequence of lattices $\Lambda,\Lambda_1,\ldots,\Lambda_L$, is nested if $\Lambda\subseteq\Lambda_{L}\subseteq\cdots\subseteq\Lambda_{1}$. 

A nested lattice codebook can be represented by a pair of nested $n$-dimensional lattice $\Lambda_c\subseteq\Lambda_f$, and is formed by the set of all points of the fine lattice $\Lambda_f$ within the Voronoi region $\mathcal{V}_c$ of the coarse lattice $\Lambda_c$, i.e., $\mathcal{C}=\Lambda_f\cap\mathcal{V}_c$. The rate of this nested lattice codebook is
\begin{IEEEeqnarray}{rCl}
R=\frac{1}{n}\log|\mathcal{C}|=\frac{1}{n}\log\frac{\mathrm{Vol}(\mathcal{V}_c)}{\mathrm{Vol}(\mathcal{V}_f)}.
\end{IEEEeqnarray}

Similarly, nested lattice codebook with various rates can be constructed by appropriately selecting pairs of lattices from a sequence of nested lattices $\Lambda\subseteq\Lambda_{L}\subseteq\cdots\subseteq\Lambda_{1}$.

\noindent We will need the following definitions \cite{CS88} in the remainder of this paper. 
\begin{definition}[MSE Goodness] A sequence of lattices $\Lambda^{(n)}\subset\mathds{R}$ is \textit{good for mean-squared error (MSE) quantization} if $\lim_{n\rightarrow \infty}G(\Lambda^{(n)})=1/2\pi e$.
\end{definition}

\begin{definition}[AWGN Goodness]
Let $\mathbf{z}$ be a length-$n$ i.i.d. Gaussian vector, $\mathbf{z}\sim\mathcal{N}(0,\sigma^2_{Z}\mathbf{I}_{n})$. The \textit{volume-to-noise ratio of a lattice} is defined as
$\mu(\Lambda,\epsilon)\triangleq\mathrm{Vol}(\mathcal{V})^{2/n}/\sigma^2_Z$
where $\sigma_Z^2$ is chosen such that $\mathrm{Pr}\{\mathbf{z}\in\mathcal{V}\}=\epsilon$. A sequence of lattices $\Lambda^{(n)}$ is \textit{good for Additive White Gaussian Noise (AWGN)} if 
$
\lim_{n\rightarrow\infty}\mu(\Lambda^{(n)},\epsilon)=2\pi e$, $\epsilon \in (0,1);
$
and, for fixed volume-to-noise ratio no smaller than $2\pi e$, i.e., $\mu(\Lambda^{(n)},\epsilon)>2\pi e$, $\mathrm{Pr}\{\mathbf{z}\notin\mathcal{V}^{(n)}\}$ 
decays exponentially in $n$.
\end{definition}

\begin{figure*}
\centering
\includegraphics[width=0.95\textwidth]{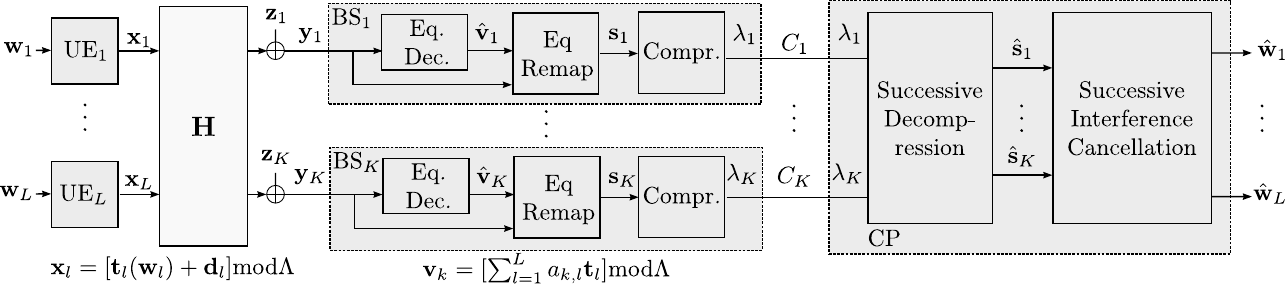}
\caption{Block diagram of the scheme Quantized-Compute-and-Forward.} \label{fig:QCoF}
\end{figure*}
\section{Quantized Compute-and-Forward}\label{sec:QCoF}

In this section, we describe our first scheme, denoted by Quantized-Compute-and-Forward (QCoF). A block diagram of this scheme is shown in Figure~\ref{fig:QCoF}. 

\subsection{Coding Scheme and Achievable Rate}\label{sec:QCoF_subsecA}

\noindent First, we summarize the main idea of the scheme QCoF, and the rationale behind it; a more detailed description will follow.  In the spirit of \cite{Nazer:IT:2011}, every BS computes a linear combination (with integer-valued coefficients) of the users' messages.
The computed equations are correlated, and forwarding each computed equation as is incurs in information redundancy at the CP. Rather, each BS first remaps the decoded equation into one that relates the users' input symbols (as opposed to the information messages).
The remapped equations at different BSs are therefore correlated. Then, the BSs sends a lossy compressed version of it to the CP in a distributed manner, taking into account this correlation. More precisely, let $\dv a_k=[a_{k1},\hdots,a_{kL}]$, $k=1,\hdots,K$, denote the integer coefficients of the equation on the users' messages $\mathbf{\hat{v}}_k$ computed at BS $k$ as done in \cite{Nazer:IT:2011}. BS $k$ uses its channel output $\dv y_k$ to map this equation into one, $\mathbf{s}_k=\mathbf{a}_k\mathbf{X}$, that relates  the input symbols from the users. (See below or \cite{Nazer:Succ:2012} for the details of this step). Because of the rate-limitation on the link that connects it to the CP, BS $k$ sends a quantized version of $\mathbf{s}_k$ to the CP. The compression is performed successively a-la  Wyner-Ziv \cite{Wyner1978}, i.e., utilizing the previously decompressed equations as side information at the CP. Let $\boldsymbol\lambda_k$ be the description of $\mathbf{s}_k$ as produced by BS $k$. The compression index associated with $\boldsymbol\lambda_k$ is sent to the CP over the pipe of capacity $C_k$. The CP collects all the compression indices and decompresses the compressed equations as $\{\mathbf{\hat{s}}_1,\ldots,\mathbf{\hat{s}}_K\}$. Then estimates the messages $\{\mathbf{w}_1,\hdots,\mathbf{w}_L\}$ successively, using a successive interference cancellation decoder. It is important to note that, as opposed to the standard CoF of \cite{Nazer:IT:2011}, due to the modified receiver, the computed equations need not be linearly independent, nor there should be as many unknown symbols as equations in general. See the comments in Section \ref{sec:QCoF_subsecB}. 

\vspace{2mm}

The following theorem provides the rate tuples achievable by QCoF for the Gaussian CRAN model of Figure~\ref{fig:Schm}.

\begin{theorem}\label{lem:QCoF}
For a set of integer-valued equation coefficients $\mathbf{A}=[\mathbf{a}_1^T,\ldots,\mathbf{a}_K^T]^T$, not necessarily full rank, the rate tuples $(R_1,\ldots,R_L)$ achievable by QCoF are given for $l=[1:L]$, $k=[1:K]$ by
\begin{IEEEeqnarray}{rCl}
R_l\leq \min_{k}\left\{ \min_{k:a_{kl}\neq0} R_{\mathrm{co}}(\mathbf{h}_k,\mathbf{a}_k,\mathrm{SNR}),\frac{1}{2}\log g_{ll}^2\right\},\nonumber
\end{IEEEeqnarray}
where the \textit{computational rate}, $R_{\mathrm{co}}(\mathbf{h},\mathbf{a},\mathrm{SNR})$, is defined as
\begin{IEEEeqnarray}{rCl}
R_{\mathrm{co}}(\mathbf{h},\mathbf{a},\mathrm{SNR})\!\triangleq\! \frac{1}{2}\log^+\left(\frac{\mathrm{SNR}}{\mathbf{a}^{T}(\mathrm{SNR}^{-1}\mathbf{I}_K+\mathbf{h}\mathbf{h}^{T})^{-1}\mathbf{a}}\right),\nonumber
\end{IEEEeqnarray}
and $g_{ll}$ are the diagonal terms of the unique lower triangular matrix $\mathbf{G}$ satisfying the Cholesky decomposition
\begin{IEEEeqnarray}{rCl}\label{eq:CholeskyQCoF}
\mathbf{GG}^T=\mathbf{I}+\mathrm{SNR}\mathbf{A}^T\mathbf{\Sigma}^{-1}\mathbf{A},
\end{IEEEeqnarray}
with\vspace{-2mm}
\begin{align}
\mathbf{\Sigma} &= \mbox{diag}(\sigma_1^{2},\hdots,\sigma_K^{2}), \\
\sigma_{k}^{2} &= \frac{ \mathbf{a}_k\!\left( \mathbf{A}_{k-1}^T\mathbf{\Sigma}^{-1}_{k-1}\mathbf{A}_{k-1}\!+\!
\mathrm{SNR}^{-1}\mathbf{I}_K \right)^{-1}\mathbf{a}_k^T\nonumber}{2^{2C_k}-1},
\end{align}
and $\mathbf{A}_k$ and $\mathbf{\Sigma}_{k}$ are respectively, the $k{\times}L$ matrix  $\mathbf{A}_k\triangleq [\mathbf{a}_1^T,\ldots,\mathbf{a}_k^T]^T$ and the $k{\times}k$ matrix $\mathbf{\Sigma}_{k}= \mbox{diag}(\sigma_1^{2},\hdots,\sigma_k^{2}) $.
\end{theorem}

\textbf{Outline of Proof:} In this proof outline, we describe the main encoding, compressing and decoding procedures in QCoF, and analyze the achievable rate tuples.  

\subsubsection{Nested lattice codebook construction}\label{sssec:BodebookConst}
At the encoders, we construct a chain of nested lattice codes using $(L+1)$ $n$-dimensional nested lattices $\Lambda\subseteq\Lambda_{L}\subseteq\cdots\subseteq\Lambda_{1}$ based on Construction A as in \cite{Nazer:IT:2011}, \cite{Erez20014Achieving}. Let $\mathbf{B}\in\mathds{R}^{n\times n}$ be the generator matrix of lattice $\Lambda$, scaled such that $\sigma^2(\Lambda)=\mathrm{SNR}$. Then, let the function $g: \mathds{F}_p\rightarrow\mathds{Z}$  denote the map between the prime-sized finite field $\mathds{F}_p$ and the corresponding subset of integers $\{0,1,2,\ldots, p-1\}$. We use $g^{-1}: \mathds{Z}\rightarrow\mathds{F}_p$ to denote its inverse. We assume that $g$ or $g^{-1} $ are applied element-wise to vectors and matrices. The fine lattices are constructed as follows. Let $\mathbf{G}\in \mathds{F}_p^{n\times k}$ be a random matrix with i.i.d. elements uniformly drawn over $\mathds{F}_p$. Let $\mathbf{G}_l$ the first $k_l$ columns of $\mathbf{G}$ (assume $k_1\geq \ldots \geq k_{L}$) and define the codebook $\mathcal{C}_l=\{\mathbf{G}_l\mathbf{c}:\mathbf{c}\in \mathds{F}_p^{k_l}\}$. 
Then, construct $\tilde{\Lambda}_{l} = p^{-1}g(\mathcal{C}_l)+\mathds{Z}^n$ and let $\Lambda_l = \mathbf{B}\tilde{\Lambda}_{l}$. These lattices are good for both AWGN and MSE \cite{Erez20014Achieving}. Also, we let $\sigma^2(\Lambda_{l})=\sigma_{l}^{2}(1+\delta_n)$, for some $\sigma_{l}^{2}>0$ whose choice will be given below,  and $\delta_n\rightarrow 0$ as $n$ increases.
We form a set of codebooks $\{\mathcal{L}_{l}\}_{l=1}^{L}$ with $\mathcal{L}_l = \Lambda_k\cap\mathcal{V}$ and rates $R_{1},\hdots,R_{L}$. The elements of $\mathcal{L}_l$ are denoted by $\{\mathbf{t}_l\}$. UE $l$, $l=1,\hdots,L$, maps $\mathbf{w}_l$ to the elements in $\mathcal{L}_l$ as $\mathbf{t}_l(\mathbf{w}_l)=\phi(\mathbf{w}_l)$ where $\phi(\mathbf{w}_l)\triangleq[\mathbf{B}p^{-1}g(\mathbf{G}\mathbf{w}_l)]\mod\Lambda_l$.  Mapping $\phi$ is invertible, and $\mathbf{w}_l$ can be recovered from $\mathbf{t}_l(\mathbf{w}_l)$ as $\mathbf{w}_l=\phi^{-1}(\mathbf{t}_l)$ \cite[Lemma 5]{Nazer:IT:2011 }. These lattice codebooks are used for the \mbox{transmission from the users.}

\noindent Similarly, let $\{(\Lambda_k^r,\Lambda_{q,k}^r)\}_{k=1}^{K}$ be $K$ pairs of $n$-dimensional nested lattices satisfying $\Lambda_k^r \subseteq \Lambda_{q,k}^r$, and forming a set of codebooks $\{\mathcal{L}_k^r\}_{k=1}^{K}$ of rates $R_1^{r},\hdots,R_K^{r}$. These nested lattice pairs are based on Construction A as above and are good for AWGN and MSE. Also, we let the second moments be chosen such that $\sigma^2(\Lambda_k^r)=\sigma_{\mathrm{eff},k}^{2,r}(1+\delta_n)$ and $\sigma^2(\Lambda_{q,k}^r)=\sigma_{\mathrm{eff},q,k}^{2,r}$, where $\sigma_{\mathrm{eff},q,k}^{2,r}$ and $\sigma_{\mathrm{eff},k}^{2,r}$ are parameters whose choices will be given below. We use the codebook $\mathcal{L}_k^r$, $k=1,\hdots,K$, for the compression at BS $k$; and we  denote its elements as $\{\boldsymbol\lambda_k\}$.

\subsubsection{User transmission at UE $l$}

At UE $l$, message $\mathbf{w}_l$, of rate $R_{l}$, is mapped into a lattice point from its codebook, $\mathbf{t}_l(\mathbf{w}_l)\in\mathcal{L}_{l}$ and the channel input is generated as
\begin{IEEEeqnarray}{rCl}\label{eq:ChanInp}
\mathbf{x}_l = [\mathbf{t}_l(\mathbf{w}_l)+\mathbf{u}_l]\mod\Lambda,
\end{IEEEeqnarray}
where $\mathbf{u}_l$ is a dither that is chosen uniformly distributed over the Voronoi region $\mathcal{V}$ of 
$\Lambda$. Note that $\mathrm{E}[|\mathbf{x}_l|^2]= n\sigma^2(\Lambda)=n\mathrm{SNR}$.

\subsubsection{Equation decoding at BS $k$}
Equation decoding at the BSs is done similarly to CoF in \cite{Nazer:IT:2011}. Let us define $\mathbf{T}\triangleq [\mathbf{t}_1^T,...,\mathbf{t}_L^T]^T$.
BS $k$ decodes an integer linear combination of the transmitted messages from the observation $\mathbf{y}_k$, defined as
$\mathbf{v}_k\triangleq \left[\mathbf{a}_{k}\mathbf{T}\right]\mathrm{mod}\;\Lambda
$, where $\mathbf{a}_k\triangleq [a_{k,1},...,a_{k,L}]\in \mathds{Z}^{L}$. 
BS $k$ scales the received signal with $\beta_k$ and computes
\begin{IEEEeqnarray}{rCl}
\mathbf{r}_{k}&\triangleq&\left[\beta_k\mathbf{y}_k-\sum_{l=1}^L a_{k,l}\mathbf{u}_l\right]\mathrm{mod }  \;\Lambda\nonumber
=\left[\mathbf{v}_k+\mathbf{z}_{\mathrm{eff},k}\right]\mathrm{mod}\;\Lambda\label{eq:decoding},
\end{IEEEeqnarray}
where the effective noise is defined as
$\mathbf{z}_{\mathrm{eff},k}\triangleq\sum_{l=1}^{L}(\beta_k h_{k,l}-a_{k,l})\mathbf{x}_l+\beta_k\mathbf{z}_k$, is statistically independent of $\mathbf{v}_k$ due to the crypto-lemma \cite{Nazer:IT:2011}, and has variance 
\begin{IEEEeqnarray}{rCl}\label{eq:effNoise}
\sigma^{2}_{\mathrm{eff},k} \triangleq \frac{1}{n}\mathrm{E}[\|\mathbf{z}_{\mathrm{eff},k}\|^2]= \mathrm{SNR}\|\beta_k \mathbf{h}_k-\mathbf{a}_k\|^2+1.
\end{IEEEeqnarray}

Let $\Lambda_{l^*(k)}$ be the finest lattice involved in equation $\mathbf{v}_k$ with coefficients $\mathbf{a}_k$, defined as $\Lambda_{l^*(k)}\triangleq \{\Lambda_l: l=\max\{l:a_{k,l}\neq 0\}\}$. Then, the BS produces an estimate for $\mathbf{v}_k$ by quantizing $\mathbf{r}_k$ with the associated lattice quantizer to $\Lambda_{l^*(k)}$, as
\begin{IEEEeqnarray}{rCl}\label{eq:EQDec}
\hat{\mathbf{v}}_k=
[Q_{\Lambda_{l^*(k)}}(\mathbf{r}_k)]\mathrm{mod}\;\Lambda.
\end{IEEEeqnarray}
An error in decoding $\mathbf{v}_k$ occurs when the effective noise lies outside the Voronoi region of $\Lambda_{l^*(k)}$. Following \cite[Theorem 5]{Nazer:IT:2011}, since $\Lambda_{l}$ are AWGN good, and $\Lambda$ is good for MSE, for sufficiently large $n$, it can be shown that $\epsilon_k\triangleq \mathrm{Pr}[\mathbf{\hat{v}}_k\neq\mathbf{v}_k]$ decays exponentially to zero in $n$, if 
\begin{IEEEeqnarray}{rCl}\label{eq:CoFCond}
R_l<\min_{k:a_{kl}\neq0} \frac{1}{2}\log\left(\frac{\mathrm{SNR}}{ \sigma^{2}_{\mathrm{eff},k} }\right).
\end{IEEEeqnarray}
The right hand side (RHS) of (\ref{eq:CoFCond}) is maximized by selecting $\beta_k$ as the minimum mean square error (MMSE) coefficient minimizing (\ref{eq:effNoise}), i.e., 
\begin{IEEEeqnarray}{rCl}
\beta_k^*= \frac{ \mathrm{SNR} \mathbf{h}_k^T\mathbf{a}_k}{ \mathrm{SNR} \|\mathbf{h}_k\|^2+1}.
\end{IEEEeqnarray}
Then, $\mathbf{v}_k$ is successfully decoded if for $l=1,\ldots,L$
\begin{IEEEeqnarray}{rCl}\label{eq:CoFCond}
R_l=\frac{1}{2}\log\left(\frac{\sigma^2(\Lambda)}{\sigma^2(\Lambda_{l})}\right) < \min_{k:a_{kl}\neq0} R_{\mathrm{co}}(\mathbf{h}_k,\mathbf{a}_k,\mathrm{SNR}).
\end{IEEEeqnarray}

\subsubsection{Equation remapping}
Given the decoded equation $\hat{\mathbf{v}}_k$ and the received signal $\mathbf{y}_k$, BS $k$ remaps its decoded  equation $\hat{\mathbf{v}}_k$ to an equation on the users' symbols with coefficients $\mathbf{a}_k$, given by  $\mathbf{s}_k=\mathbf{a}_k\mathbf{X}$, by calculating
\begin{IEEEeqnarray}{rCl}
\boldsymbol\phi_k&=&[\hat{\mathbf{v}}_k+\sum_{l=1}^{L}a_{k,l}\mathbf{u}_l]\mod \Lambda, \quad\text{and}\\
\mathbf{s}_k&=&Q_{\Lambda}(\beta^*_k\mathbf{y}_k-\boldsymbol\phi_k)+\boldsymbol\phi_k.
\end{IEEEeqnarray}
For sufficiently large $n$, the probability of having an error in remapping, i.e., $\mathrm{Pr}\{\mathbf{s}_k\neq \mathbf{a}_k\mathbf{X}\}$ is arbitrarily small, provided $ R_{\mathrm{co}}(\mathbf{h}_k,\mathbf{a}_k,\mathrm{SNR})>0$ \cite{Nazer:Succ:2012}.

\subsubsection{Lossy compression at BS $k$}
BS $k$ compresses $\mathbf{s}_k$ at rate $C_k$ and forwards the compression codeword $\boldsymbol\lambda_k$ to the CP as follows.
Signal $\mathbf{s}_k$ is added a dither $\mathbf{u}^{\mathrm{r}}_k$, uniformly distributed over the Voronoi region of $\Lambda_k^r$, $\mathcal{V}^r_k$, and quantized to the nearest fine lattice point in $\Lambda^r_{q,k}$. Then modulo reduction over the coarse lattice $\Lambda^r_{k}$ is applied to obtain $\boldsymbol\lambda_k$: 
 \begin{IEEEeqnarray}{rCl}
  \boldsymbol\lambda_k &=& Q_{\Lambda^r_{q,k}} ( \mathbf{s}_k + \mathbf{u}_k^{r}) \mod \Lambda^r_k,\\
  &=&( \mathbf{s}_k + \mathbf{u}_k^{r}-\mathbf{z}^r_{\mathrm{eq},k}) \mod \Lambda^r_k,
 \end{IEEEeqnarray}
where $\mathbf{z}^r_{\mathrm{eq},k}\triangleq ( \mathbf{s}_k + \mathbf{u}_k^{r})\mod \Lambda^r_{q,k}$ is the quantization noise with variance $\sigma^2(\Lambda^r_k)$, and it is uniformly distributed over $\mathcal{V}^r_{q,k}$ and independent of $\mathbf{s}_k$.

BS $k$ forwards the codeword $\boldsymbol\lambda_k\in\mathcal{L}^r_k$ to the CP over the finite-capacity link. This transmission is successful as long as the rate of $\mathcal{L}^r_k$ is below the rate of the link, i.e., 
  \begin{equation}\label{eq:LinkRate}
 C_k \geq  \frac{1}{n}  \log \left( \dfrac{\mathrm{Vol}(\Lambda^r_k)}{\mathrm{Vol}(\Lambda^r_{q,k})} \right)=\frac{1}{2}  \log \left( \dfrac{\sigma^2(\Lambda^r_k)}{\sigma(\Lambda^r_{q,k})} \right).
  \end{equation} 

Inequality (\ref{eq:LinkRate}) is satisfied with the choice, justified below,  of the following parameters $\sigma^2(\Lambda_{k}^r)=\sigma_{\mathrm{eff},k}^{2,r}(1+\delta_n)$ and $\sigma(\Lambda^r_{q,k})=\sigma_{\mathrm{eff},q,k}^{2,r}$, and $\delta_n'\rightarrow 0$, $\delta_n'>1/2\log(1+\delta_n)$, where
\begin{IEEEeqnarray}{rCl}
\sigma_{\mathrm{eff},k}^{2,r} &=& \nu(\mathbf{s}_k|\hat{\mathbf{s}}_1^{k-1})+\sigma^2(\Lambda^r_{q,k}),\label{eq:Sigmaqr}\\
\sigma_{\mathrm{eff},q,k}^{2,r} &=& \nu(\mathbf{s}_k|\hat{\mathbf{s}}_1^{k-1})/(2^{ (C_{k}-\delta_n')}-1),\label{eq:Sigmar} 
\end{IEEEeqnarray}
where 
\begin{IEEEeqnarray}{rCl}
\nu(\mathbf{s}_k|\hat{\mathbf{s}}_1^{k-1})\triangleq \mathbf{a}_k\!\left(
\mathbf{A}_{k-1}^T\mathbf{\Sigma}^{-1}_{k-1}\mathbf{A}_{k-1}\!+\!
\mathrm{SNR}^{-1}\mathbf{I}_K
\right)^{-1}\mathbf{a}_k^T,\nonumber
\end{IEEEeqnarray}
 is the MMSE obtained if $\mathbf{s}_k$ is estimated using a linear estimator from the $k-1$ reconstructions $\hat{\mathbf{s}}_1^{k-1}$ after decompression, as detailed below.

\subsubsection{Successive decompression at CP}

After receiving the compression codewords $(\boldsymbol\lambda_1,...,\boldsymbol\lambda_K)$, the CP successively reconstructs $\{\mathbf{s}_1,\ldots,\mathbf{s}_K\}$ as $\{\hat{\mathbf{s}}_1,\ldots, \mathbf{\hat{s}}_K\}$, starting from $\mathbf{s}_1$\footnote{Note that the reconstruction order can be optimized. However, we consider a fixed decoding order for simplicity. Any other order can be considered by relabeling the BSs.} 
The CP decompresses $\mathbf{s}_k$ as follows. Assume that $k-1$ sequences $\mathbf{\hat{s}}_1^{k-1}\triangleq[\mathbf{\hat{s}}_1^T,...,\mathbf{\hat{s}}_{k-1}^T]^T$ have already been reconstructed. As shown below, the successfully reconstructed  $\mathbf{\hat{s}}_k$ is equivalent to $\mathbf{\hat{s}}_k = \mathbf{s}_k+\mathbf{z}^{r}_{\mathrm{eq},k}$. Then, the CP computes an effective side information sequence $\tilde{\mathbf{s}}_k$ with a linear estimate of $\mathbf{s}_k$ with coefficients $\boldsymbol\gamma_k\in \mathds{R}^{1\times k-1}$ as
\begin{IEEEeqnarray}{rCl}\label{eq:MMSEestimationQCoF}
\tilde{\mathbf{s}}_k=\boldsymbol{\gamma}_k\hat{\mathbf{s}}_1^{k-1}=\boldsymbol{\gamma}_k(\mathbf{A}_{k-1}\mathbf{X}+\mathbf{Z}^r_{\mathrm{eq},k-1}),
\end{IEEEeqnarray}
where $\mathbf{Z}^r_{\mathrm{eq},k} = [\mathbf{z}^{r,T}_{\mathrm{eq},1},\ldots,\mathbf{z}^{r,T}_{\mathrm{eq},k}]^T$ has variance $\mathbf{\Sigma}_{k}$.

The CP reconstructs $\mathbf{s}_k$ with $\boldsymbol\lambda_k$ and the effective side information sequence $\mathbf{\tilde{s}}_k$, by computing: 
\begin{IEEEeqnarray}{rCl}
\mathbf{\hat{s}}_k
 &=&  [\boldsymbol\lambda_k - \mathbf{u}^{r}_k -  \mathbf{\tilde{s}}_k]\!\! \!  \mod  \Lambda_{k}^r + \mathbf{\tilde{s}}_k \nonumber\\
& =&  [ \mathbf{s}_k -  \mathbf{z}^r_{\mathrm{eq},k}  -   \tilde{\mathbf{s}}_k ]\!\! \! \mod  \Lambda^r_k + \tilde{\mathbf{s}}_k \nonumber\\
 &\overset{(\mathrm{c.d.})}{=}&    \mathbf{s}_k +  \mathbf{z}^r_{\mathrm{eq},k},\label{eq:AddnoiseQ}
 \end{IEEEeqnarray}
where $(\mathrm{c.d.})$ holds if decompression is correct. A decompression error occurs when $\bar{\mathbf{s}}_k\triangleq (\mathbf{s}_k -\tilde{\mathbf{s}}_k)+  \mathbf{z}^r_{\mathrm{eq},k}$, lies outside the Voronoi region of $\Lambda^r_k$, i.e., 
$\mathrm{Pr}\{\hat{\mathbf{s}}_k\neq  \mathbf{s}_k +  \mathbf{z}^r_{\mathrm{eq},k}\}=\mathrm{Pr}\{\bar{\mathbf{s}}_k\notin \mathcal{V}(\Lambda^r_k)\}$. From \cite[Lemma 8]{Nazer:IT:2011}, the density of $\bar{\mathbf{s}}_k$ can be upper bounded (times a constant) by the density of an i.i.d. zero mean Gaussian vector $\bar{\mathbf{s}}_k^*$ whose variance $\sigma^{2,r}_{\mathrm{eff},k}$ approaches that of $\bar{\mathbf{s}}_k$ as $n\rightarrow \infty$. Since $\Lambda^r_k$ is AWGN good, the probability of error $\epsilon^r_k\triangleq\mathrm{Pr}\{\bar{\mathbf{s}}_k^*\notin \mathcal{V}(\Lambda^r_k)\}$ decays to zero exponentially in $n$, as long as the volume to noise ratio satisfies 
$\mu(\Lambda^r_k,\epsilon^r_k)>2\pi e$. If this occurs, the  probability of decompression error $\mathrm{Pr}\{\bar{\mathbf{s}}_k\notin \mathcal{V}(\Lambda^r_k)\}$ also decays to zero exponentially in $n$.
From the definition of the normalized second moment,  we have $\mu(\Lambda^r_k,\epsilon^r_k)>2\pi e$ if
\begin{eqnarray}
\sigma^{2} (\Lambda_k^r) > 2\pi e G(\Lambda_k^r) \sigma^{2,r}_{\mathrm{eff},k}.
\end{eqnarray}
Since $\Lambda_k^r$ are good for MSE quantization, i.e., $G(\Lambda_k^r)\rightarrow 1/2\pi e$, for sufficiently large $n$ decompression is successful since $\sigma^{2} (\Lambda_k^r)=(1+\delta_n)\sigma^{2,r}_{\mathrm{eff},k}$, and we have
 \begin{IEEEeqnarray}{rCl}
\sigma^{2,r}_{\mathrm{eff},k} &= & \dfrac{1}{n} \mathds{E}  ||  (\mathbf{s}_k -\tilde{\mathbf{s}}_k )+  \mathbf{z}^r_{\mathrm{eq},k}||^2 \\
&=& \dfrac{1}{n} \mathds{E}  ||  \mathbf{s}_k - \tilde{\mathbf{s}}_k||^2 + \sigma^2(\Lambda^r_{q,k}),\label{eq:dither}  \\
&=&\mathrm{SNR}\|\mathbf{a}_k-\boldsymbol\gamma_k\mathbf{A}_k\|^2+\boldsymbol\gamma_k\mathbf{\Sigma}_{k-1}\boldsymbol\gamma_k^T+ \sigma^2(\Lambda^r_{q,k})\label{eq:dither2}  ,
\end{IEEEeqnarray}
where (\ref{eq:dither}) follows since $\mathbf{z}^r_{\mathrm{eq},k}$ is independent of $\mathbf{s}_k$ and $\tilde{\mathbf{s}}_k$ and since $1/n\mathds{E}[\|\mathbf{z}_{\mathrm{eq}^r}\|^2]=\sigma^2(\Lambda_{k}^r)$;(\ref{eq:dither2}) follows since $\mathbf{Z}^r_{\mathrm{eq},k-1}$ and $\mathbf{X}$ are independent, and since $1/n\mathds{E}[\mathbf{X}\mathbf{X}^T]=\mathrm{SNR}\mathbf{I}$.

Substituting (\ref{eq:dither2}) in (\ref{eq:LinkRate}) we observe that the range of feasible $\sigma^2(\Lambda^r_{q,k})$ in (\ref{eq:feasibleSigma}) is enlarged by choosing $\boldsymbol\gamma_k$ in ( \ref{eq:MMSEestimationQCoF}) as the optimal linear MMSE estimator, i.e., 
\begin{IEEEeqnarray}{rCl}
\nu(\mathbf{s}_k|\hat{\mathbf{s}}_1^{k-1})=
\min_{\boldsymbol\gamma}\mathrm{SNR}\|\mathbf{a}_k-\boldsymbol\gamma_k\mathbf{A}_k\|^2+\boldsymbol\gamma_k\mathbf{\Sigma}_{k-1}\boldsymbol\gamma_k^T,
\end{IEEEeqnarray} 
which can be found as
\begin{IEEEeqnarray}{rCl}
\boldsymbol{\gamma}_k^* = \mathrm{SNR}\mathbf{a}_k\mathbf{A}_{k-1}^T(\mathrm{SNR}\mathbf{A}_{k-1}\mathbf{A}_{k-1}^T+\mathbf{I}_{k-1}+\boldsymbol\Sigma_{k-1})^{-1}\nonumber.
\end{IEEEeqnarray}
Then, (\ref{eq:LinkRate})  is given as
\begin{IEEEeqnarray}{rCl}\label{eq:feasibleSigma}
C_k\geq \frac{1}{2}\log\left(\frac{ \nu(\mathbf{s}_k|\hat{\mathbf{s}}_1^{k-1}) }{\sigma^2(\Lambda^r_{q,k}) } +1\right)\!+\!\frac{1}{2}\log(1+\delta_{n}).
\end{IEEEeqnarray}
Finally, $\hat{\mathbf{s}}_1,\ldots,\mathbf{\hat{s}}_K$ are successfully decompressed at the CP since (\ref{eq:feasibleSigma}) holds for $k=1,\ldots,K$ due to the choice (\ref{eq:Sigmar}).


\subsubsection{Decoding at the CP with successive interference cancellation} The $K$ decompressed sequences $\{\hat{\mathbf{s}}_1,\ldots, \mathbf{\hat{s}}_K\}$ can be modeled as
$\hat{\mathbf{S}}=\mathbf{AX}+\mathbf{Z}^r_{\textrm{eff},K}$. Note that the channel noise $\mathbf{Z}$ has been removed, i.e., the signal has been denoised.
 
The CP applies successive interference cancellation in order to recover all the transmitted messages \cite{Cioffi1995, Ordentlich2013}. First, the CP performs linear MMSE estimation of $\mathbf{X}$ from $\hat{\mathbf{S}}$ as $\mathbf{\hat{X}}=\mathbf{\Gamma}\hat{\mathbf{S}}$, using the optimal linear MMSE filter 
\begin{IEEEeqnarray}{rCl}
\mathbf{\Gamma} = \mathrm{SNR}\mathbf{A}(\mathrm{SNR}\mathbf{AA}^T+\mathbf{\Sigma})^{-1}.
\end{IEEEeqnarray}
Then, for the MMSE error $\mathbf{E}\triangleq (\mathbf{\Gamma}\mathbf{A}-\mathbf{I})\mathbf{X}+\mathbf{B}\mathbf{Z}^r_{\textrm{eff}}$, the CP computes the unique Cholesky decomposition of its covariance  matrix, $\mathbf{K}_E = \mathrm{SNR}(\mathbf{GG}^T)^{-1}$, where 
$\mathbf{GG}^T$ is given as in (\ref{eq:CholeskyQCoF}),
and $\mathbf{G}$ is a lower triangular matrix with strictly positive diagonal entries. The estimated symbols are distributed as
\begin{IEEEeqnarray}{rCl}
\mathbf{\hat{X}}&=&\mathbf{X}+\sqrt{\mathrm{SNR}}\mathbf{G}^{-1}\mathbf{N}\label{eq:MMSEnoise},
\end{IEEEeqnarray}
where \label{eq:MMSEnoise} $\mathbf{N}$ is an equivalent white noise with covariance $\mathbf{I}$.

The CP decodes the messages $\{\mathbf{w}_1,\ldots,\mathbf{w}_L\}$ successively,  by decoding $\{\mathbf{t}_1,\ldots,\mathbf{t}_L\}$ starting from $\mathbf{t}_1$. To decode $\mathbf{t}_1$, the CP gets the estimation from the first column of $\mathbf{\Gamma}$, $\mathbf{\hat{x}}_{1}=\boldsymbol{\gamma}^{(1)}\hat{\mathbf{S}}$, and computes 
\begin{IEEEeqnarray}{rCl}
\mathbf{\hat{t}}_1&=& Q_{\Lambda_1}([\boldsymbol\gamma^{(1)}\hat{\mathbf{S}}+\mathbf{u}_1]\mod\Lambda)\\
&=& Q_{\Lambda_1}([\mathbf{t}_1+\sqrt{\mathrm{SNR}}g_{11}^{-1}\mathbf{n}_1]\mod\Lambda).
\end{IEEEeqnarray}
Similarly to (\ref{eq:EQDec}), a decoding error occurs if the effective noise lies outside the Voronoi region of $\Lambda_{1}$. Since the effective noise observed satisfies $\sigma_{\mathrm{eff},1}^{2,\mathrm{sic}}=1/n\mathrm{E}[ \|\sqrt{\mathrm{SNR}}g_{11}^{-1}\mathbf{n}_1\|^2] =  \mathrm{SNR}(g_{11}^2)^{-1}$, the probability of error in decoding decays to zero exponentially in $n$ if 
\begin{IEEEeqnarray}{rCl}
R_1\leq\frac{1}{2} \log\left(\frac{\mathrm{SNR}}{\sigma_{\mathrm{eff},1}^{2,\mathrm{sic}}}\right)=\frac{1}{2}\log g_{11}^2.
\end{IEEEeqnarray}
If $\mathbf{t}_1$ is successfully decoded, the CP estimates $\mathbf{n}_1$ as
\begin{IEEEeqnarray}{rCl}
\hat{\mathbf{n}}_1 = \frac{1}{g_{11}\sqrt{\mathrm{SNR}}}\left[\boldsymbol\gamma^{(1)}\hat{\mathbf{S}}-\mathbf{t}_1\right]\mod \Lambda\overset{(\mathrm{c.d.})}{=}\mathbf{n}_1.
\end{IEEEeqnarray}
where (c.d.) holds if there is no estimation error. 
An error in estimating the effective noise, i.e., $\mathrm{Pr}[\mathbf{n}_1\neq \mathbf{\hat{n}}_1]$ occurs if  $\mathbf{n}_1$ lies outside the Voronoi region of  $\Lambda$. Note that successful estimation occurs with high probability for large $n$ since $g_{11}\sqrt{\mathrm{SNR}}\mathbf{n}_1\in\mathcal{V}_k\subseteq\mathcal{V}$. Then, the CP uses $\hat{\mathbf{n}}_1$ to reduce the noise and recovers $\mathbf{t}_2$ similarly to $\mathbf{t}_1$ by applying
\begin{IEEEeqnarray}{rCl}
\mathbf{t}_2 &=& Q_{\Lambda_2}\left([\boldsymbol\gamma^{(2)}\mathbf{\hat{S}}-\sqrt{\mathrm{SNR}}\mathbf{g}_1\mathbf{\hat{n}}_1+\mathbf{u_2}]\mod\Lambda\right)\\
&=& Q_{\Lambda_2}([\mathbf{t}_2+\sqrt{\mathrm{SNR}}g_{22}^{-1}\mathbf{n}_2]\mod\Lambda).
\end{IEEEeqnarray}
Iterating this process, since the effective noise observed for each lattice codeword $\mathbf{t}_l$ is $\sigma_{\mathrm{eff},l}^{2,\mathrm{sic}}= \mathrm{SNR}(g_{ll}^2)^{-1}$, each  $\mathbf{t}_l$  can be decoded successively provided 
\begin{IEEEeqnarray}{rCl}\label{eq:SICCond}
R_l\leq\frac{1}{2} \log\left(\frac{\mathrm{SNR}}{\sigma_{\mathrm{eff},l}^{2,\mathrm{sic}}}\right)=\frac{1}{2}\log g_{ll}^2.
\end{IEEEeqnarray}

Once $\{\mathbf{t}_1,\ldots,\mathbf{t}_L\}$ have been decoded, the CP recovers the messages $\{\mathbf{w}_1,\ldots,\mathbf{w}_L\}$ as $\mathbf{w}_l = \phi^{-1}(\mathbf{t}_l)$.

Finally,  by the union bound, it follows that for sufficiently large $n$, the probability of error $\mathrm{Pr}\{(\hat{\mathbf{w}}_1,...,\hat{\mathbf{w}}_L)\neq(\mathbf{w}_1,...,\mathbf{w}_L)\}$ vanishes if (\ref{eq:CoFCond}) and (\ref{eq:SICCond}) are satisfied. Note that since $R_l=\frac{1}{2}\log\left(\sigma^2(\Lambda)/\sigma^2(\Lambda_{l})\right)$ and $\sigma^2(\Lambda)=\mathrm{SNR}$, both conditions are satisfied if the second moment of the fine lattices $\Lambda_l$, $\sigma^2(\Lambda_l)$, are chosen such that 
\begin{IEEEeqnarray}{rCl}
\sigma_{l}^{2}=\max\{
\max_{k:a_{kl}\neq0}\sigma^{2}_{\mathrm{eff},k},\sigma_{\mathrm{eff},l}^{2,\mathrm{sic}}  \}.
\end{IEEEeqnarray}
This completes the proof of Theorem \ref{lem:QCoF}.\qed

%
%
%
%
%

%
%
%

\subsection{Comments}\label{sec:QCoF_subsecB}
The scheme QCoF shares some elements with the standard CoF of~\cite{Nazer:IT:2011}. Also, it has some connections at high level with successive Wyner-Ziv \cite{Somekh:2007:IT,DelCoso:2009:TWir, Sanderovich:2009:IT, Hwan:2013:VT, ZhouYu:2013:JSAC,Park:2013:SPLett }. The following remarks help better understanding these connections.

\begin{remark}
In the scheme of \cite{Tan2016Assym}, every relay node computes an equation that relates the users's information messages and sends it to the CP. The correlation among the computed equations is exploited there by performing distributed lossless compression. (Note that this is possible in \cite{Tan2016Assym} because the computed equations lie in discrete sets). In our case, by opposition to \cite{Tan2016Assym}, the computed equations relate the users' input signals, rather than the messages; and so lie each in a continuum. For this reason, they are sent compressed \textit{in a lossy manner} to the CP.  This allows us to trade-off appropriately the desired compression level (related to the allowed backhaul capacity) and the quality of signals recovery at the CP. \qed
\end{remark}
\begin{remark}\label{rem:CoFVareq}
At the CP, the users' messages are not obtained by directly inverting the linear system of equations, but from the effective output formed by the lossy compressed equations forwarded to the CP. 
Therefore, the decoded equations do not need to be linearly independent, nor there need to be as many of them as users.
Hence, while QCoF applies in general, it seems more suited to situations in which there are more users than relay nodes, i.e., $L > K$. Note that, in this case, i.e., if $L > K$, a variation of the original CoF of \cite{Nazer:IT:2011} can still be applied, e.g., by appropriately selecting a subset of the relay which will be required to compute several equations each (ensuring that $L$ linearly independent equations are computed in total). However, this generally results in some performance degradation (see Remark \ref{rem:SuccessiveMinima}).\qed
\end{remark}

\begin{remark}
The procedure of computing, at each relay node, an equation $\mathbf{s}_k$ that relates the users' input symbols (instead of one that relates the users' information messages) can, at a high level, be seen as some form of (partial) \textit{denoising} of the received signal at that relay node prior to the compression. Compared to the successive Wyner-Ziv of \cite{Somekh:2007:IT,DelCoso:2009:TWir, Sanderovich:2009:IT, Hwan:2013:VT, ZhouYu:2013:JSAC,Park:2013:SPLett }, the advantage here is in removing out part of the channel noise before the compression, so as to save some compression rate.
 Alternatively, it can be seen as inducing more correlation among the signals that are to be compressed distributedly.\qed
\end{remark}

\subsection{Sum-Rate Optimization and Equation Selection for QCoF}\label{sec:QCoF_subsecC}

In what follows,  we consider the problem of maximization of the sum-rate allowed by QCoF. Using Theorem \ref{lem:QCoF}, the sum-rate optimization can be formulated as
\begin{subequations}\label{eq:SumRateCond}
\begin{IEEEeqnarray}{rCl}
R^{\mathrm{QCoF}}_{\mathrm{sum}}&\triangleq& \max_{\mathbf{A},\mathbf{r}}\sum_{l=1}^Lr_l\label{eq:SumRateCond0QCoF}\\
&&\text{s.t. } r_l\leq \min_{k:a_{kl}\neq0} R_{\mathrm{co}}(\mathbf{h}_k,\mathbf{a}_k,\mathrm{SNR}), \label{eq:SumRateCond1QCoF}\\
&&\quad\;\; r_l\leq \frac{1}{2}\log g_{ll}^2, \quad  l\in[1:L],  k\in [1:K], \label{eq:SumRateCond2QCoF}\\
&&\quad \;\;\mathbf{A}\in\mathds{Z}^{K\times L }, \mathbf{r}\in\mathds{R}^L_+\label{eq:SumRateCond3QCoF}
\end{IEEEeqnarray}
\end{subequations} 
where $\{g_{ll}\}$ are the diagonal elements of the matrix $\mathbf
{G}$ satisfying \eqref{eq:CholeskyQCoF}.
 
The optimization problem (\ref{eq:SumRateCond0QCoF})-(\ref{eq:SumRateCond3QCoF})  is a mixed integer non-linear problem, which is in general difficult to solve using standard techniques. 
 In order to select the integer-coefficients that maximize $R^{\mathrm{QCoF}}_{\mathrm{sum}}$, one can consider either of the following two possible solutions.
\subsubsection{Exhaustive search} In this case, the search space can by limited by considering only those rows of $\mathbf{A}$ for which the RHS of (\ref{eq:SumRateCond1QCoF}) is non-zero. This leads to integer-coefficients which satisfy $\|\mathbf{a}_k\|^2\leq 1+\mathrm{SNR}\|\mathbf{h}_k\|^2.$
In this manner, the complexity of exhaustive search can be reduced. However, it remains generally prohibitive, especially for high SNR values.

\subsubsection{Approximate solution}

Note that the RHS of \eqref{eq:SumRateCond2QCoF} as well depends on the selected integer-coefficients vector $\mathbf{a}_k$ (implicitly, through the diagonal elements $\{g_{ll}\}$). The joint optimization of the RHS of \eqref{eq:SumRateCond1QCoF} and the RHS of \eqref{eq:SumRateCond2QCoF} is not easy in general. In this method, we select $\{\mathbf{a}_k\}$ so as to maximize the RHS of \eqref{eq:SumRateCond1QCoF} only (and then evaluate the RHS of \eqref{eq:SumRateCond2QCoF} using the found set of integer coefficients). This can be performed using the well known Lenstra-Lenstra-Lov\'az (LLL) algorithm \cite{HongCaire:IT:2013} as follows. Let \mbox{$\mathbf{F}_k\mathbf{F}_k^{T} = \left(\mathbf{I}_L+\mathrm{SNR}\mathbf{h}_k\mathbf{h}_k^{{T}}\right)^{-1} $} be a Cholesky decomposition. The computational rate in (\ref{eq:SumRateCond1QCoF}) can be written as
\begin{IEEEeqnarray}{rCl}\label{eq:RCO_1}
R_{\mathrm{co}}(\mathbf{h}_k,\mathbf{a}_k,\mathrm{SNR}) = -\frac{1}{2}\log^+ \| \mathbf{F}_k^{T}\mathbf{a}_k\|^2.
\end{IEEEeqnarray}
The RHS of (\ref{eq:RCO_1}) is maximized by finding $\mathbf{a}_k$ as the shortest lattice point of the $L$ dimensional lattice spanned by $\mathbf{F}_k^T$ as follows. Apply the LLL  algorithm $\mathbf{F}_k^{T}$ to find the reduced matrix $\mathbf{F}_k^{' T}$ and compute $\mathbf{\tilde{A}}_k = \mathbf{F}_k^{' T}(\mathbf{F}_k^{T})^{-1}$. Then, choose $\mathbf{a}_k$ as the row of $\mathbf{\tilde{A}}_k$ with the smallest norm.

\begin{remark}\label{rem:SuccessiveMinima}
Recalling that the allowed transmission rate is constrained by the smallest computation rate, it is apparent that the more equations a relay node needs to compute the smaller the allowed transmission rate. 
Alternatively, this can be seen by noting that, at the relay node $k$, $k=1,\hdots,K$, each computed equation corresponds to one of the successive minima of the lattice expanded by $\mathbf{F}_k^T$.
\end{remark}

\section{Jointly Quantized Compute-and-Forward}\label{sec:JQCoF}

In this section, we describe our second scheme for the  model of Figure~\ref{fig:Schm}. This scheme is a generalization of QCoF and we denote it as Jointly-Quantized-Compute-and-Forward (JQCoF). 

\subsection{Coding Scheme and Achievable Sum-Rate}\label{sec:JQCoF_subsecA}
First, we summarize briefly the main ideas of the scheme JQCoF. Like in QCoF, each BS computes an equation that relates the users' symbols. However, by opposition to the scheme QCoF, here each BS compresses not only the computed equation but also its output signal. This is performed using multivariate distributed compression. More specifically, BS $k$, $k=1,\hdots,K$, obtains $\mathbf{s}_k=\mathbf{a}_k\mathbf{X}$ as in QCoF. Then, it compresses jointly the vector $\boldsymbol\theta_k=[\mathbf{s}_k^T,\mathbf{y}_k^T]^T$, where $\mathbf{y}_k$ denotes its received signal, in the spirit of the successive multivariate Wyner-Ziv compression of \cite{DelCoso:2009:TWir, Hwan:2013:VT}. That is, at BS $k$, the vector sequence $\boldsymbol\theta_k$ is linearly combined into $\mathbf{r}_k$ (see below for more details on this step). Then, the $\mathbf{r}_{k,1}$ and $\mathbf{r}_{k,2}$ are quantized independently. Let $\boldsymbol\lambda_{k,1}$ and $\boldsymbol\lambda_{k,2}$ be the description of $\mathbf{r}_{k,1}$ and $\mathbf{r}_{k,2}$, respectively, as produced by BS $k$. The available rate $C_k$ of the error-free link is allocated between the two descriptions and the corresponding indices sent to the CP. The CP collects all the indices and reconstructs the compressed signals successively as $\{\mathbf{\hat{s}}_1,\mathbf{\hat{y}}_1,\ldots,\mathbf{\hat{s}}_k,\mathbf{\hat{y}}_k\}$, utilizing at each decompression step the signals already decompressed as side information. Finally, the users' messages $\{\mathbf{w}_1,\ldots,\mathbf{w}_L\}$ are decoded successively  with a successive interference cancellation decoder, in a way that is essentially similar to with the scheme QCoF.

\begin{remark}\label{rem:Tradeoff}
The joint multivariate compression only considers the innovation of $\mathbf{y}_k$ w.r.t. $\mathbf{s}_k$. Using this joint compression, the scheme JQCoF can trade-off appropriately between the denoising capabilities of QCoF and exploiting the correlation among the channel outputs at the relay nodes as in the successive Wyner-Ziv of \cite{Hwan:2013:VT}. In particular, this allows to cluster the relay nodes in such a way that a subset of them do not decode any equation (if this is not helpful) and simply perform Wyner-Ziv compression and the remaining nodes apply QCoF. For example, in the specific case in which no equations are decoded at any BS, the performance of JQCoF reduces to the performance to that of single antenna SWZ \cite{ Hwan:2013:VT}. \qed
\end{remark}

\begin{remark}\label{rem:RemainingSignal}
Given that $\mathbf{s}_k$ can be seen as a denoised version of the output $\mathbf{y}_k$, the reader may wonder why having the same relay note sending lossy descriptions of both signals may still be beneficial in general (from a sum-rate viewpoint). To see this, consider, for example, the case in which the channel coefficients are integer-valued. In this case, it is clear that the best denoised equation is one with the same coefficients as the channel, and the signal would then be fully denoised. In the more realistic case in which the channel coefficients are not integer-valued, part of the output signal  is not captured by the computed equation.  For this reason,  it is in general beneficial to also convey a description of the innovation of $\mathbf{y}_k$ w.r.t. $\mathbf{s}_k$ to the CP using multivariate compression. \qed
\end{remark}

The following theorem provides the rate tuples achievable by JQCoF for the Gaussian CRAN model of Figure~\ref{fig:Schm}.
\begin{theorem}\label{lem:JQCoF}
For a set of integer-valued equation coefficients $\mathbf{A}=[\mathbf{a}_1^T,\ldots, \mathbf{a}_K^T]^T$, not necessarily full rank, and an arbitrarily small $\epsilon>0$, the rate tuples achievable by JQCoF are given for $l=[1:L]$, $k=[1:K]$ by
\begin{IEEEeqnarray}{rCl}
R_l\leq \min_{k}\left\{ \min_{k:a_{kl}\neq0} R_{\mathrm{co}}(\mathbf{h}_k,\mathbf{a}_k,\mathrm{SNR}),\frac{1}{2}\log g_{ll}^2\right\},\nonumber\\
\text{s.t. }C_k \geq \frac{1}{2}\log(1+\eta_{k,1}\lambda_{k,1})+\frac{1}{2}\log(1+\eta_{k,2}\lambda_{k,2}), \nonumber
\end{IEEEeqnarray}
where $g_{ll}$ are the diagonal terms of triangular matrix $\mathbf{G}$ from the unique Cholesky decomposition
\begin{IEEEeqnarray}{rCl}\label{eq:CholeskyJQCoF}
\mathbf{GG}^T=\mathbf{I}+\mathrm{SNR}\mathbf{\bar{H}}_K^T(\mathbf{I}_{2K}+\bar{\boldsymbol
\Omega}_K)^{-1}\mathbf{\bar{H}}_K
\end{IEEEeqnarray}
where $\bar{\mathbf{H}}_k\!=\![\mathbf{\bar{h}}_{1}^T,\ldots,\mathbf{\bar{h}}_{k}^T]^T$, $\mathbf{\bar{h}}_{k}=\mathbf{C}_{\mathbf{n}}^{-1/2}[\mathbf{a}_k^T,\mathbf{h}_k^T]^T$; $\mathbf{C}_{\mathbf{n}} =[\epsilon,0;0,1] $  and $ \bar{\boldsymbol\Omega}_k = \text{diag}( \boldsymbol\Omega_{1},\ldots, \boldsymbol\Omega_{k})$, with
\begin{IEEEeqnarray}{rCl} \label{eq:CovStuct}
\boldsymbol\Omega_k=\mathbf{U}_k\text{diag}(\eta_{k,1},\eta_{k,2})^{-1}\mathbf{U}_k^T,
\end{IEEEeqnarray}
and where $\mathbf{U}_k$ and $\mathbf{D}_k=\text{diag}(\lambda_{k,1},\lambda_{k,2})$  follow from the the singular value decomposition (SVD) of matrix
\begin{IEEEeqnarray}{rCl}
\mathbf{K}_{\boldsymbol\theta_k|\hat{\boldsymbol\theta}_1^{k-1}}&=&\mathbf{\bar{h}}_{k-1}\mathbf{K}_{\mathbf{x}|\hat{\boldsymbol\theta}_1^{k-1}}\mathbf{\bar{h}}_{k-1}^T+\mathbf{I}_2=\mathbf{U}_k\mathbf{D}_k\mathbf{U}_k^T\label{eq:CovTheta},
\end{IEEEeqnarray}
where
\begin{IEEEeqnarray}{rCl}
\mathbf{K}_{\mathbf{x}|\hat{\boldsymbol\theta}_{1}^{k-1}}=(\mathrm{SNR}^{-1}\mathbf{I}+\bar{\mathbf{H}}_{k-1}^T(\mathbf{I}_{2(k-1)}+\bar{\boldsymbol\Omega}_{k-1})^{-1}\bar{\mathbf{H}}_{k-1})^{-1},\nonumber
\end{IEEEeqnarray}
\end{theorem} 

\textbf{Outline of Proof:} For reasons of brevity, we describe only  the steps in which JQCoF differs from QCoF. The remaining steps are similar to in the previous section.

\subsubsection{Nested lattice codebook construction}
As in QCoF, for transmission from the UEs we construct a chain of nested lattice codes using $(L+1)$ $n$-dimensional nested lattices $\Lambda\subseteq\Lambda_{L}\subseteq\cdots\subseteq\Lambda_{1}$ based on Construction A and the corresponding set of nested codebooks $\{\mathcal{L}_{l}\}_{l=1}^{L}$ with $\mathcal{L}_l = \Lambda_k\cap\mathcal{V}$ and rates $R_{1},\hdots,R_{L}$ . We assume second moments $\sigma^2(\Lambda)=\mathrm{SNR}$ and $\sigma^2(\Lambda_{l})=\sigma_{l}^{2}(1+\delta_n)$, for some $\sigma_{l}^{2}>0$ whose choice will be given below, and $\delta_n\rightarrow 0$\mbox{ as $n$ increases. }

For compression at the BS we generated two compression codebooks at each BS. We consider two sets of $K$ pairs of $n$-dimensional lattices $\{(\Lambda_{k,1}^{r},\Lambda_{q,k,j}^{r})\}_{k=1}^K$ for $j=1,2$,  such that $\Lambda_{k,j}^{r} \subseteq \Lambda_{q,k,j}^{r}$, $k=1,\ldots,K$ and $j=1,2$, forming two sets of codebooks $\{\mathcal{L}_{k,j}^r\}_{k=1}^{K}$, $j=1,2$, of rates $R_{k,j}^{r}$. Due to its construction, $\Lambda_{k,j}^{r}$ $\Lambda^{r}_{q,k,j}$ are good for MSE and good for AWGN. We let the second moment  be chosen such that
\begin{IEEEeqnarray}{rCl}\label{eq:sigmas-JQCoF}
\sigma^2(\Lambda_{k,j}^{r})&=&(1+\delta_n)(\lambda_{k,j}+\eta_{k,j}^{-1})\quad \text{and}\quad\\
\sigma^2(\Lambda_{q,k,j}^{r})&=&\eta_{k,j}^{-1}.
\end{IEEEeqnarray}
The choice of this parameters is justified below. We denote the elements in codebook $\{\mathcal{L}_{k,j}^r\}$ by $\{\boldsymbol\lambda_{k,j}\}$.

\subsubsection{User transmission at UE $l$}
UE $l$ maps message $\mathbf{w}_l$ into the channel input $\mathbf{x}_l$ as in (\ref{eq:ChanInp}).

\subsubsection{Equation decoding at BS $k$}
At BS $k$ lattice equation $\mathbf{v}_k$ with integer coefficients $\mathbf{a}_k$ is decoded as in Section \ref{sec:QCoF}. Thus decoding is successful provided (\ref{eq:CoFCond}) holds.

\subsubsection{Equation remapping}

At BS $k$, $\mathbf{\hat{v}}_k$ is remapped to $\mathbf{s}_k$ as in Section \ref{sec:QCoF}, and is successful under the same conditions.

\subsubsection{Joint Compression at BS $k$}
The BS $k$ compresses the two signals $\boldsymbol\theta_k\triangleq[\mathbf{s}^T_k,\mathbf{y}_k^T]^T$. First, it generates an i.i.d. Gaussin noise sequence with variance $\epsilon\geq 0$, i.e.,  $\mathbf{\tilde{n}}_k\sim \mathcal{N}(0,\epsilon)$, applies the following linear transform\footnote{The addition of the Gaussian noise $\mathbf{\tilde{n}}_k$ and whitenning through $\mathbf{C_n}^{-\frac{1}{2}}$ is to relate the problem to the multivariate SWZ model studied in \cite{DelCoso:2009:TWir, Hwan:2013:VT}, as discussed later. However, in general this is not required.} 
\begin{IEEEeqnarray}{rCl}\label{eq:LinearTrans}
\mathbf{r}_k=\mathbf{U}^T_k\mathbf{C}_{\mathbf{n}}^{-1/2}[\mathbf{s}_k^T+\mathbf{\tilde{n}}_k,\mathbf{y}_k^T]^T,\label{eq:UncTrans}
\end{IEEEeqnarray}
Then, BS $k$  compresses each component $\mathbf{r}_{k,j}$  by computing, for $j=1,2$:
\begin{IEEEeqnarray}{rCl}
\boldsymbol\lambda_{k,j} &=& Q_{\Lambda^{r}_{q,k,j}} ( \mathbf{r}_{k,j} + \mathbf{u}_{k,j}^{r}) \mod \Lambda^{r}_{k,j},\\
  &=&( \mathbf{r}_{k,j} + \mathbf{u}_{k,j}^{r}-\mathbf{z}^r_{\mathrm{eq},k,j}) \mod \Lambda^r_{k,j},
\end{IEEEeqnarray}
where $\mathbf{z}^r_{k,j}\triangleq ( \mathbf{r}_{k,j} + \mathbf{u}_{k,j}^{r})\mod \Lambda^r_{q,k,j},
$ is the quantization noise with variance $\sigma^2(\Lambda_{q,k,j}^{r})$, uniformly distributed over the Voronoi region of $\Lambda^r_{q,k,j}$ and independent of $\mathbf{r}_{k,j}$.

Then, the  BS $k$ forwards the codewords $\boldsymbol\lambda_{k,1},\boldsymbol\lambda_{k,2}$ to the CP over the finite-capacity link. Note that the rate $C_k$ has to be shared between the two descriptions such that
 \begin{IEEEeqnarray}{rCl}
 C_k &\geq& \frac{1}{2}\log \left( \dfrac{\sigma^2(\Lambda^{r}_{k,1})}{\sigma^2(\Lambda^{r}_{q,k,1})} \right)+\frac{1}{2}\log \left( \dfrac{\sigma^2(\Lambda^{r}_{k,2})}{\sigma^2(\Lambda^{r}_{q,k,2})} \right)\nonumber\\
 &=& \frac{1}{2}\log(1+\eta_{k,1}\lambda_{k,1})+\frac{1}{2}\log(1+\eta_{k,2}\lambda_{k,2})
+\log(1+\delta_n).\label{eq:CondJQCoF2}
  \end{IEEEeqnarray}

\subsubsection{Successive decompression at CP}

After receiving the compression codewords $(\boldsymbol\lambda_{1,j},...,\boldsymbol\lambda_{K,j})$, $j=1,2$, the CP successively reconstructs the transformed components  $\{\mathbf{r}_1,\ldots,\mathbf{r}_1\}$ as $\{\hat{\mathbf{r}}_1,\ldots,\hat{\mathbf{r}}_K\}$ starting from $\mathbf{\hat{r}}_1$. To reconstruct $\mathbf{\hat{r}}_k$, the signals already reconstructed $\{\mathbf{\hat{r}},\ldots,\mathbf{\hat{r}}_{k-1}\}$ are used as side information available at the CP. Then, each $\hat{\boldsymbol\theta}_k$ is reconstructed from $\mathbf{\hat{r}}_k$ as $\hat{\boldsymbol\theta}_k = \mathbf{U}_k\mathbf{\tilde{r}}_k$. 
As shown below, the reconstructed signals $\mathbf{r}_k$ and $\hat{\boldsymbol\theta}_k$ can be modeled as
\begin{IEEEeqnarray}{rCl}
\mathbf{\hat{r}}_k& =& \mathbf{U}^T_k\bar{\mathbf{h}}_k\mathbf{X}+\mathbf{\bar{Z}}_k+\mathbf{Z}^{r}_{k},\\
\hat{\boldsymbol\theta}_k&=& \bar{\mathbf{h}}_k\mathbf{X}+\mathbf{\bar{Z}}_k+\mathbf{Z}^{r,\mathrm{eff}}_{k} \label{eq:thetarec},
\end{IEEEeqnarray}
where $\mathbf{\bar{Z}}_k\triangleq\mathbf{C}_{\mathbf{n}}^{-1/2}[\mathbf{\tilde{n}}_k^T,\mathbf{z}_k^T]^T$, is a whitened noise which has covariance matrix $\mathbf{I}_{2}$, $\mathbf{Z}^{\mathrm{r}}_{k}=[\mathbf{z}^{r,T}_{k,1},\mathbf{z}^{r,T}_{k,2}]^T$, is the quantization noise and has covariance matrix $\text{diag}(\eta_{k,1},\eta_{k,2})^{-1}$; and  $\mathbf{Z}^{\mathrm{r,\mathrm{eff}}}_{k}=\mathbf{U}_k[\mathbf{z}^{r,T}_{k,1},\mathbf{z}^{r,T}_{k,2}]^T$, is the transformed quantization noise, which has covariance matrix $\boldsymbol\Omega_k$ as in (\ref{eq:CovStuct}).

To decompresses $\mathbf{\hat{r}}_k$, the  CP computes the effective side information $\mathbf{\tilde{r}}_{k}$ by linearly combining the $k-1$ decompressed sequences $\mathbf{\hat{r}}_1^{k-1}\triangleq[\mathbf{\hat{r}}_{1}^T,...,\mathbf{\hat{r}}_{k-1}^T]^T$ with $\boldsymbol{\gamma}_{k}\in \mathds{R}^{2\times k-1}$ as
\begin{IEEEeqnarray}{rCl}\label{eq:MMSEestimation}
\tilde{\mathbf{r}}_{k}=\boldsymbol{\gamma}_{k}\hat{\mathbf{r}}_1^{k-1},
\end{IEEEeqnarray}
We chose $\boldsymbol{\gamma}_{k}$ to be the linear MMSE estimator of $\mathbf{r}_{k}(t)$ given $\mathbf{\hat{r}}_1^{k-1}(t)$, $t=1,\ldots, n$, given by
\begin{IEEEeqnarray}{rCl}
\boldsymbol{\gamma}_{k}= \mathrm{SNR}\mathbf{U}_{k}\mathbf{\bar{h}}_k\mathbf{\bar{H}}_{k-1}^T\mathbf{\bar{U}}_{k-1}^T\boldsymbol\Sigma_{\hat{r},k}^{-1},
\end{IEEEeqnarray}
where 
\begin{IEEEeqnarray}{rCl}
\boldsymbol\Sigma_{\hat{r},k}&\triangleq& 
\mathbf{\bar{U}}_{k-1}^T(\mathrm{SNR}\mathbf{\bar{H}}_{k-1}\mathbf{\bar{H}}_{k-1}^T+\mathbf{I}_{2(k-1)}+\bar{\boldsymbol\Omega}_k)\mathbf{\bar{U}}_{k-1}\nonumber.
\end{IEEEeqnarray}
Note that due to the orthonormality of the matrix of eigenvectors $\mathbf{U}_k$, we have
\begin{IEEEeqnarray}{rCl}\label{eq:CovMatU}
\mathds{E} [ (\mathbf{r}_{k}(t) -\tilde{\mathbf{r}}_{k}(t) )
(\mathbf{r}_{k}(t) -\tilde{\mathbf{r}}_{k}(t) )^T]=\mathbf{U}_k^T \mathbf{K}_{\boldsymbol\theta_k|\hat{\boldsymbol\theta}_{1}^{k-1}}\mathbf{U}_k. 
\end{IEEEeqnarray}
where $\mathbf{K}_{\boldsymbol\theta_k|\hat{\boldsymbol\theta}_{1}^{k-1}}$ is defined as in (\ref{eq:CovTheta}) and corresponds to the MMSE error matrix of estimating $\boldsymbol\theta_k$ from $\{\hat{\boldsymbol\theta}_1,\ldots,\hat{\boldsymbol\theta}_{k-1}\}$ with a linear MMSE estimator, similarly to (\ref{eq:MMSEestimation}).

Therefore, the decompression of component $j$ of $\mathbf{r}_k$ is done, similarly to  (\ref{eq:AddnoiseQ}), as follows
\begin{IEEEeqnarray}{rCl}
\mathbf{\hat{r}}_{k,j}
 &=&  (\boldsymbol\lambda_{k,j} - \mathbf{u}^{r}_{k,j} -  \mathbf{\tilde{r}}_{k,j})\!\! \!  \mod  \Lambda_{k,j}^{r} + \mathbf{\tilde{r}}_{k,j} \nonumber\\
 &\overset{(\mathrm{c.d.})}{=}&    \mathbf{r}_{k,j} +  \mathbf{z}^{r}_{k,j},\label{eq:cd2}
  \end{IEEEeqnarray}
where equality $(\mathrm{c.d.})$ in (\ref{eq:cd2}) holds as long as decompression is successful.  
Similarly to (\ref{eq:AddnoiseQ}),  the probability of decompression error decays exponentially to zero in $n$ if 
\begin{eqnarray}\label{eq:condJQCoF}
\sigma^{2} (\Lambda_{k,j}^r) > 2\pi e G(\Lambda_{k,j}^r) \sigma^{2,r}_{\mathrm{eff},k,j},
\end{eqnarray}
where $\sigma^{2,r}_{\mathrm{eff},k,j}$ is the variance of an 
i.i.d. zero mean Gaussian vector whose variance $\sigma^{2,r}_{\mathrm{eff},k}$ approaches that of $(\mathbf{r}_{k,j} -\tilde{\mathbf{r}}_{k,j} )+  \mathbf{z}^{r}_{k,j}$ as $n\rightarrow \infty$. Note that (\ref{eq:condJQCoF}) holds due to (\ref{eq:sigmas-JQCoF}), since 
\begin{IEEEeqnarray}{rCl}
\sigma^{2,r}_{\mathrm{eff},k,j} &= & 
 \dfrac{1}{n} \mathds{E}  ||  (\mathbf{r}_{k,j} -\tilde{\mathbf{r}}_{k,j} )+  \mathbf{z}^{r}_{k,j}||^2\\
&=& \dfrac{1}{n} \mathds{E}  ||  (\mathbf{r}_{k,j} -\tilde{\mathbf{r}}_{k,j} )||^2+  \dfrac{1}{n} \mathds{E}  || \mathbf{z}^{r}_{k,j}||^2\label{eq:dither11}\\
&=&[\mathbf{U}_k^T\mathbf{K}_{\boldsymbol\theta_k|\hat{\boldsymbol\theta}_1^{k-1}}\mathbf{U}_k]_{jj}+\sigma^2(\Lambda^{r}_{q,k,j})\label{eq:dither22}\\
&=&[\mathbf{D}_k]_{jj}+\sigma^2(\Lambda^{r}_{q,k,j})\label{eq:dither223}\\
&=&\lambda_{k,j}+\eta_k^{-1},
\end{IEEEeqnarray}
where (\ref{eq:dither11}) follows since $\mathbf{z}^r_{k,j}$ is independent of $\mathbf{r}_{k,j}$ and  $\tilde{\mathbf{r}}_{k,j}$; (\ref{eq:dither22}) follows from (\ref{eq:CovMatU}); (\ref{eq:dither223}) follows from (\ref{eq:CovTheta}) and since $\mathbf{U}_k^T\mathbf{U}_k=\mathbf{I}$ due to the orthonormality of the eigenvectors. Note that by transforming $\mathbf{r}_k$ as in (\ref{eq:UncTrans}), the error covariance matrix $\mathbf{K}_{\boldsymbol\theta_k|\hat{\boldsymbol\theta}_{1}^{k-1}}$ is diagonalized. See Remark \ref{rem:Omega} below.


%

\subsubsection{Decoding at the CP with successive interference cancellation} After successively decompressing the $K$ signals $\{\hat{\boldsymbol\theta}_1,\ldots, \hat{\boldsymbol\theta}_K\}$ the CP applies successive interference cancellation similarly to QCoF to recover $\{\mathbf{w}_1,\ldots,\mathbf{w}_L\}$, starting form $\mathbf{w}_1$.
First, the CP applied the linear MMSE estimator of $\mathbf{X}$ from given $\{\hat{\boldsymbol\theta}_1,\ldots, \hat{\boldsymbol\theta}_K\}$, as $\mathbf{\hat{X}}=\mathbf{\Gamma}[\hat{\boldsymbol\theta}_1^T,\ldots, \hat{\boldsymbol\theta}_K^T]^T$, where $\boldsymbol\Gamma$ is given by the filter 
\begin{IEEEeqnarray}{rCl}
\mathbf{\Gamma} = \mathrm{SNR}\mathbf{\bar{H}}_K(\mathrm{SNR}\mathbf{\bar{H}}_K\mathbf{\bar{H}}_K^T+\mathbf{I}_{2K}+\boldsymbol\Omega_{K})^{-1}.\nonumber
\end{IEEEeqnarray}
The CP computes the unique Cholesky decomposition of the MMSE error matrix, 
$\mathbf{K}_E = \mathrm{SNR}(\mathbf{GG}^T)^{-1}$, where 
$\mathbf{GG}^T$ is given as in (\ref{eq:CholeskyJQCoF}).
and $\mathbf{G}$ is a lower triangular matrix with strictly positive diagonal entries. 
Applying the successive interference cancellation decoding as in QCoF,  each  $\mathbf{t}_l$ and associated message  $\mathbf{w}_l$ can be successively decoded provided 
\begin{IEEEeqnarray}{rCl}\label{eq:CondJQCoF3}
R_l\leq\frac{1}{2} \log\left(\frac{\mathrm{SNR}}{\sigma_{\mathrm{eff},l}^{2,\mathrm{sic}}}\right)=\frac{1}{2}\log g_{ll}^2,
\end{IEEEeqnarray}
where the effective noise observed for each lattice codeword $\mathbf{t}_l$ is found as $\sigma_{\mathrm{eff},l}^{2,\mathrm{sic}}= \mathrm{SNR}(g_{ll}^2)^{-1}$. Then, the CP recovers the users messages from $\{\mathbf{t}_1,\ldots,\mathbf{t}_L\}$. 

Finally, by the union bound, it follows that for sufficiently large $n$ the probability of error $\mathrm{Pr}\{(\hat{\mathbf{w}}_1,...,\hat{\mathbf{w}}_L)\neq(\mathbf{w}_1,...,\mathbf{w}_L)\}$ can be made arbitrarily small provided (\ref{eq:CoFCond}), (\ref{eq:CondJQCoF2})  and (\ref{eq:CondJQCoF3}), are satisfied. Note that  similarly to QCoF, (\ref{eq:CoFCond}) and (\ref{eq:CondJQCoF3}) are satisfied if
 the second moment of the fine lattices $\Lambda_l$, $\sigma^2(\Lambda_l)$, are chosen with 
\begin{IEEEeqnarray}{rCl}
\sigma_{l}^{2}=\max\{
\max_{k:a_{kl}\neq0}\sigma^{2}_{\mathrm{eff},k},\sigma_{\mathrm{eff},l}^{2,\mathrm{sic}}  \}.
\end{IEEEeqnarray}
This completes the proof of Theorem \ref{lem:JQCoF}.
\qed

\begin{remark}\label{rem:Omega}
The design of $\boldsymbol\Omega_k$ in (\ref{eq:CovStuct}) is chosen to have the same eigenvector structure, $\mathbf{U}_k$, as the covariance noise that maximizes the achievable sum-rate of multivariate Successive Wyner Ziv utilizing Gaussian test channels as described in \cite{DelCoso:2009:TWir} and  \cite{Hwan:2013:VT}. Similarly to our derivation, the covariance is designed such that in their setup, the MMSE error matrix $\mathbf{K}_{\boldsymbol\theta_k|\hat{\boldsymbol\theta}_{1}^{k-1}}$ is also diagonalized as in \eqref{eq:dither223}. \qed
\end{remark}

\subsection{Sum-Rate Optimization and Equation Selection for JQCoF}\label{sec:JQCoF_subsecC}

In this section, we consider the maximization of the sum-rate achievable by JQCoF. Using Theorem \ref{lem:JQCoF}, the sum-rate optimization can be formulated as
\begin{subequations}\label{eq:SumRateCondJQoF}
\begin{IEEEeqnarray}{rCl}
R^{\mathrm{JQCoF}}_{\mathrm{sum}}&\triangleq& \max_{\mathbf{A},\boldsymbol\eta_1,\boldsymbol\eta_2,\mathbf{r}}\sum_{l=1}^Lr_l \label{eq:SumRateCond0JQCoF}\\
\text{s.t. }&& r_l\leq \min_{k:a_{kl}\neq0} R_{\mathrm{co}}(\mathbf{h}_k,\mathbf{a}_k,\mathrm{SNR}), \label{eq:SumRateCond1JQCoF}\\
&&r_l\leq \frac{1}{2}\log g_{ll}^2, \qquad   l\in[1:L],  k\in [1:K], \label{eq:SumRateCond2JQCoF}\\
&&C_k \geq \sum_{j=1}^{2}\frac{1}{2}\log(1+\eta_{k,j}\lambda_{k,j}), \label{eq:SumRateCond3JQCoF}\\
&&\mathbf{A}\in\mathds{Z}^{K\times L },
\boldsymbol\eta_{1}\in\mathds{R}^K_+,
\boldsymbol\eta_{2}\in\mathds{R}^K_+,\mathbf{r}\in\mathds{R}^L_+.\label{eq:SumRateCond4JQCoF}
\end{IEEEeqnarray}
\end{subequations} 
where $\{g_{ll}\}$ are the diagonal elements of the matrix $\mathbf
{G}$ satisfying \eqref{eq:CholeskyJQCoF}
and $\{\lambda_{1,k},\lambda_{2,k}\}$ are the eigenvalues obtained from the SVD of matrix \eqref{eq:CovTheta}.

As in QCoF, the optimization problem (\ref{eq:SumRateCond0JQCoF})-(\ref{eq:SumRateCond4JQCoF}) is a mixed integer non-linear problem, and hence it is hard to optimize. On can consider the following possible approaches to optimize the integer coefficients and the quantization noises that optimize $R^{\mathrm{JQCoF}}_{\mathrm{sum}}$.
\subsubsection{Exhaustive search}
Similarly to QCoF, we can consider exhaustive search over $\mathbf{A}$ for which the RHS in (\ref{eq:SumRateCond1JQCoF}) is non-zero, i.e., $\mathbf{a}_k$ satisfies
$\|\mathbf{a}_k\|^2\leq 1+\mathrm{SNR}\|\mathbf{h}_k\|^2$. Then, for each fixed $\mathbf{A}$, (\ref{eq:SumRateCond0JQCoF})-(\ref{eq:SumRateCond3JQCoF}) is a convex problem which can be efficiently solved. 

\subsubsection{Approximate equations}
Similarly to QCoF, we can select the coefficients $\mathbf{a}_k$ that maximizes the RHS of (\ref{eq:SumRateCond1JQCoF}). Each $\mathbf{a}_k$ can be obtained utilizing the LLL algorithm as in Section \ref{sec:QCoF_subsecC}. For such $\mathbf{A}$, (\ref{eq:SumRateCond0JQCoF})-(\ref{eq:SumRateCond4JQCoF}) is solved as a convex optimization problem. This method ignores the effect of $\mathbf{A}$ in the compression and centralized decoding captured by (\ref{eq:SumRateCond3JQCoF}). 
\subsubsection{Approximate solution for the quantization noise}
As noted, for fixed equations $\mathbf{A}$, (\ref{eq:SumRateCond0JQCoF})-(\ref{eq:SumRateCond3JQCoF}) is a convex problem. Here, we propose an approximate solution to the quantizaiton noises $(\boldsymbol\eta_{1},\boldsymbol\eta_{2})$ by solving the relaxed sum-rate optimization problem in which constraints (\ref{eq:SumRateCond2JQCoF}) are removed, based on the following observation. At BS $k$, after remapping $\mathbf{v}_k$ to $\mathbf{s}_k$ the observed signal to compress is $(\mathbf{a}_k\mathbf{X},\mathbf{y}_k)$. Thus, the problem is that of the sum-rate optimization for multi-variate successive Wyner Ziv problem studied in \cite{DelCoso:2009:TWir} and  \cite{Hwan:2013:VT}, if $(\mathbf{a}_k\mathbf{X},\mathbf{y}_k)$ were Gaussian. Next proposition shows that the same noise allocation as in \cite{Hwan:2013:VT} can be utilized.
\begin{proposition}\label{prop:Noisealloc}
A feasible solution for the quantization noises in the sum-rate problem is (\ref{eq:SumRateCond0JQCoF})-(\ref{eq:SumRateCond4JQCoF}) given by
\begin{IEEEeqnarray}{rCl}
\eta_{k,j} = \left(\frac{1}{\mu}\left(1-\frac{1}{\lambda_{k,j}}\right)-1\right)^+\label{eq:etaalloc},
\end{IEEEeqnarray}
with $\mu>0$ chosen to satisfy
\begin{IEEEeqnarray}{rCl}
\frac{1}{2}\log(1+\eta_{k,1}\lambda_{k,1})+\frac{1}{2}\log(1+\eta_{k,2}\lambda_{k,2})=C_k.
\end{IEEEeqnarray}
\end{proposition}

\begin{proof} We show that, although $(\mathbf{a}_k\mathbf{X},\mathbf{y}_k)$ are not Gaussian distributed, the  sum-rate optimization problem (\ref{eq:SumRateCond0JQCoF}) and (\ref{eq:SumRateCond2JQCoF})-(\ref{eq:SumRateCond4JQCoF}), coincides with that in \cite[Equation (6)]{Hwan:2013:VT}. To see this, let us consider an i.i.d. zero mean Gaussian variable $\mathbf{X}'\sim\mathcal{N}(0,\mathrm{SNR}\mathbf{I}_L)$, $K$ sequences $\mathbf{Y}_k'=\bar{\mathbf{h}}_k\mathbf{X}'+\mathbf{Z}_k'$, $k=1,\ldots,K$, where $\mathbf{Z}'_k\sim\mathcal{N}(0,\mathbf{I}_{2})$ is an i.i.d. zero mean Gaussian random vector, and $K$ sequences $\mathbf{\hat{Y}}_k'=\mathbf{Y}_k+\mathbf{\hat{Z}}_k'$, $k=1,\ldots,K$, where $\mathbf{Z}'\sim\mathcal{N}(0,\mathbf{\Omega}_k)$, representing the compressed signals.
For this model, the covariance matrix of the MMSE estimation error of estimating $\mathbf{Y}_k$ given $\mathbf{Y}_1^{k-1}$ is equal to $\mathbf{K}_{\boldsymbol\theta_k|\hat{\boldsymbol\theta}_1^{k-1}}$ in (\ref{eq:CovTheta}). Then, let us define $\mathbf{\hat{Y}}'=[\mathbf{\hat{Y}}_1^T,\ldots,\mathbf{\hat{Y}}_K^T]$. We have
\begin{IEEEeqnarray}{rCl}
I(\mathbf{X};\mathbf{\hat{Y}}')&=&\frac{1}{2}\log|\mathbf{I}+\mathrm{SNR}\mathbf{\bar{H}}_K^T(\mathbf{I}_{2K}+\boldsymbol\Omega_{K})^{-1}\mathbf{\bar{H}}_K|\\
&=&\frac{1}{2}\log|\mathbf{GG}^T|=
\sum_{l=1}^L\frac{1}{2}\log g_{ll}^2
\end{IEEEeqnarray}
and
\begin{IEEEeqnarray}{rCl}
I(\mathbf{Y}_k;\mathbf{\hat{Y}}_k|\mathbf{\hat{Y}}_1^{k-1})&=&
\log |\mathbf{K}_{\boldsymbol\theta_k|\hat{\boldsymbol\theta}_1^{k-1}}+\mathbf{\Omega}_k|-\log|\mathbf{\Omega}_k|\nonumber\\
&=&\sum_{j=1}^{2}\frac{1}{2}\log(1+\eta_{k,j}\lambda_{k,j}).
\end{IEEEeqnarray}
Then, it is easy to see that the optimization problem 
(\ref{eq:SumRateCond0JQCoF}) and (\ref{eq:SumRateCond2JQCoF})-(\ref{eq:SumRateCond4JQCoF})  can be written as given in \cite[Equation (6)]{Hwan:2013:VT}.
\end{proof}

\section{Numerical Results}\label{sec:WynerModel}

In this section, we provide some numerical examples that illustrate the average sum-rates obtained using QCoF and JQCoF. We consider a CRAN network with $L=3$ users and $K=2$ BSs and  channel coefficients distributed as $h_{l,k}\sim\mathcal{N}(0,1)$. We average the achievable sum-rates over 2000 channel realizations. 
We also consider the SWZ of \cite{Hwan:2013:VT}  and a variation of the CoF of \cite{Nazer:IT:2011}, in which one of the two relays decodes two equations. The schemes are compared among them, and also to the following cut-set upper bound
\begin{IEEEeqnarray}{rCl}
R^{\mathrm{CS}}_{\mathrm{sum}}=\min\left\{\frac{1}{2}\log\det(\mathbf{I}+\mathrm{SNR}\mathbf{HH}^T),\sum_{k=1}^KC_k\right\}.\nonumber
\end{IEEEeqnarray}

\begin{figure}[!t]
\begin{minipage}[t]{0.5\linewidth}
    \centering
	\includegraphics[width=0.99\textwidth]{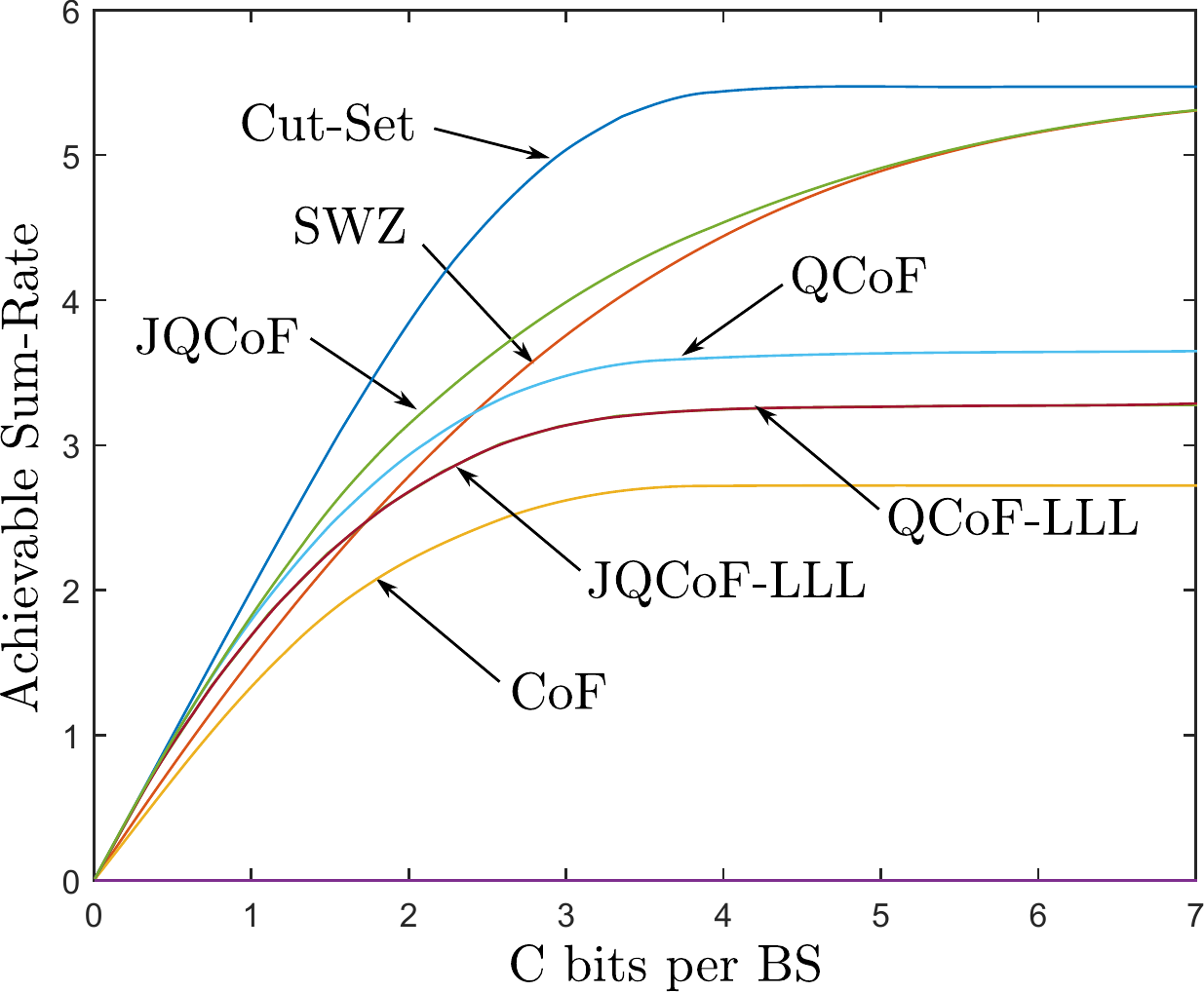}
	\vspace{-5mm}	
	\caption{Average upper and lower bounds on the  sum-rate 	for the proposed schemes with respect to  $C$ for $\mathrm{SNR}= 		5\mathrm{dB}$.}
	 \label{fig:K2_L3_SNR5_vsC}
\end{minipage}
\hspace{0.1cm}
\begin{minipage}[t]{0.5\linewidth} 
    \centering
\includegraphics[width=0.99\textwidth]{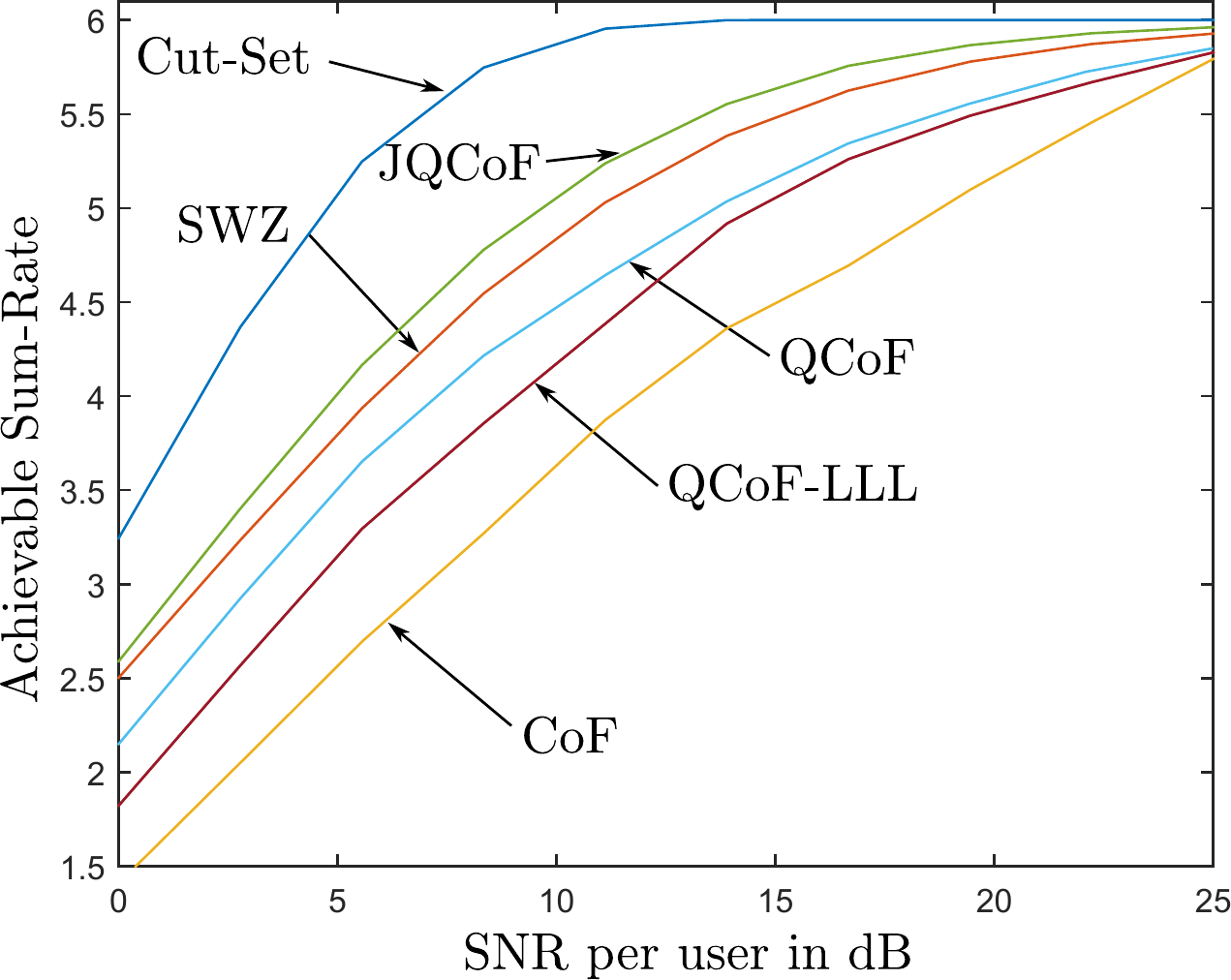}
\vspace{-5mm}	
\caption{Average upper and lower bounds on thee sum-rate for the proposed schemes with respect to the $\mathrm{SNR}$ for $C=3$.} \label{fig:RateSNR}
\end{minipage}        
\end{figure}


\noindent Figure \ref{fig:K2_L3_SNR5_vsC} depicts the evolution of the sum-rate as a function of the backhaul capacity $C$ (in this case, $C_1=\cdots=C_K = C$), for $\mathrm{SNR}=5\mathrm{dB}$. As it can be seen from the figure, our scheme QCoF outperforms CoF for all $C$ values, since it requires less equation computations, and SWZ for moderate  $C$ values. The scheme JQCoF performs better than all the other schemes. This is line with Remark \ref{rem:Tradeoff}, since JQCoF can balance its performance between QCoF and SWZ. It is seen that the sum-rate of both QCoF and CoF saturates as the backhaul capacity $C$ increases, while for SWZ and JQCoF the sum-rate approaches the cut-set bound. This follows since for large $C$ values, the  compression noise becomes negligible and the CP can decode as if the signals at the BSs were available to it. However, the performance of CoF and QCoF is limited since part of the signal is not extracted at each BS by computing the equations, as discussed in Remark \ref{rem:RemainingSignal}.  Figure \ref{fig:K2_L3_SNR5_vsC} also shows the sum-rate of the suboptimal implementations of QCoF and JQCoF, denoted by QCoF-LLL and JQCoF-LLL respectively, in which the integer coefficients are found using the LLL algorithm as explained above. Interestingly, both QCoF and JQCoF have the same performance. However, while the scheme QCoF-LLL achieves a performance close to that of QCoF, this is not the case for JQCoF-LLL and JQCoF.  

Figure \ref{fig:RateSNR} shows the  sum-rate for the proposed schemes and its suboptimal versions based on LLL with respect to the available $\mathrm{SNR}$ per user, for $C = 3$. Our scheme QCoF outperforms CoF for all $\mathrm{SNR}$ values and JQCoF achieves the best performance among the considered schemes.

\begin{appendices}

\end{appendices}

\bibliographystyle{ieeetran}
\bibliography{ref}
\newpage

\end{document}